\documentclass[submit]{smj}
\usepackage[latin9]{inputenc}
\usepackage{array}
\usepackage{booktabs}
\usepackage{amsmath}
\usepackage{graphicx}

\makeatletter

\providecommand{\tabularnewline}{\\}

\makeatother

\begin{document}
\Title{Sparse and Smooth Prior for Bayesian Linear Regression with Application to ETEX Data}
\TitleRunning{Sparse and Smooth Prior for Bayesian Linear Regression }

\Author{Luk\'a\v{s} Ulrych, V\'aclav \v{S}m\'\i{}dl}
\AuthorRunning{L. Ulrych, V. \v{S}m\'\i{}dl}
\Affiliations{
\item Institute of Information Theory and Automation, Pod vodarenskou vezi 4, Prague, Czech Republic,
}
\CorrAddress{V\'aclav \v{S}m\'\i{}dl,  Institute of Information Theory and Automation, 
Pod vodarenskou vezi 4, Prague, Czech Republic} \CorrEmail{smidl@utia.cas.cz} \CorrPhone{(+420)\;266\;952\;420} \CorrFax{(+420)\;286\;890\;378} 

\Abstract{Sparsity of the solution of a linear regression model is a common requirement, and many prior distributions have been designed for this purpose. A combination of the sparsity requirement with smoothness of the solution is also common in application, however, with considerably fewer existing prior models. In this paper, we compare two prior structures, the Bayesian fused lasso (BFL) and least-squares with adaptive prior covariance matrix (LS-APC). Since only variational solution was published for the latter, we derive a Gibbs sampling algorithm for its inference and Bayesian model selection. The method is designed for high dimensional problems, therefore, we discuss numerical issues associated with evaluation of the posterior. In simulation, we show that the LS-APC prior achieves results comparable to that of the Bayesian Fused Lasso for piecewise constant parameter and outperforms the BFL for parameters of more general shapes. Another advantage of the LS-APC priors is revealed in real application to estimation of the release profile of the European Tracer Experiment (ETEX). Specifically, the LS-APC model provides more conservative uncertainty bounds when the regressor matrix is not informative.}

\maketitle{}

\section{Introduction}

The standard linear regression model, $y=X\beta$, has probably the
largest literature on the choice of priors for the vector of parameters
$\beta$. Majority of the results was derived for the variable selections
problem \cite{george1993variable} where shrinkage priors play a key
role. However, sparsity of the solution is not the only prior knowledge
in many practical applications, such as estimation of the source term
of an atmospheric release of a pollutant \cite{stohl2012xenon}. An
equally important role in this field has the assumption of smoothness
which corresponds to the natural assumption that the release is a
piecewise continuous function. While Bayesian approaches for this
particular application exists, e.g. \cite{ganesan2014characterization,henne2016validation},
the practice is still dominated by by penalized likelihood approaches
combining penalization terms for smoothness and sparsity in L2 norm
\cite{eckhardt2008estimation} or L1 norm \cite{tibshirani2005sparsity}.
The reason is that they provide a computationally efficient solution
for high-dimensional problems. 

Relation between penalized likelihood methods and Bayesian priors
has been extensively studied especially for the Lasso problem \cite{park2008bayesian,alhamzawi2012bayesian}
and many methods inspired by this research proposed e.g. \cite{rockova2015spike-and-slab}.
However, much less work has been done on combination of the sparsity
and smoothness knowledge as presented in the fused lasso \cite{tibshirani2005sparsity}.
The Bayesian prior yielding the fused lasso approach was presented
by \cite{Kyung_BFL} where a more flexible parametrization was proposed.
An alternative hierarchical prior was proposed by \cite{Tichy2016gmd}
in tandem with Variational Bayesian inference algorithm in the context
of atmospheric modeling. This prior is not directly related to the
fused lasso formulation, since it was derived from the L2 approach
of \cite{eckhardt2008estimation}. The proposed prior is closely related
to time-space priors \cite{cai2012bayesian}, with also uses a linear
model for correlations. However, we extend this approach by another
hidden variable of unknown correlation coefficient with shrinkage
prior \cite{Tichy2016gmd}.

In this paper, derive a Gibbs sampling algorithm for smoothness and
sparsity prior for inference and Bayesian model selection. The model
selection is next used to select which of the predefined measurement
covariance matrix is appropriate. The proposed algorithm is compared
with the Variational Bayes solution of the same models and with the
Bayesian fused lasso. Both extensive simulation studies as well as
comparison on real data from the European Tracer Experiment (ETEX)
is presented.

\section{Smoothness and Sparsity Prior of the LS-APC Model}

One of the first models of sparsity is the hierarchical prior based
on Normal-Gamma models \cite{tipping2001sparse}
\begin{align}
\beta_{i} & \sim\mathcal{N}(0,\tau_{i}^{-1}), & \tau_{i} & \sim\mathcal{G}(a,b),i=1,\ldots,p,\label{eq:ard}
\end{align}
where prior distribution of the precision parameter $\tau_{i}$ is
assumed to have fixed scalar parameters $a,b$. Their typical choice
is motivated by non-infor\-ma\-ti\-ve\-ness, i.e. both of them
are very low (close to numerical precision) to approach the Jeffrey's
prior. Inference of the model with this prior has the effect of shrinking
the posterior probability of $\beta_{i}$ to zero. The same model
can be used to describe smoothness (or more exactly, piecewise smoothness)
by promoting sparsity of the derivative , i.e.
\begin{align}
\beta_{i+1}-\beta_{i} & \sim\mathcal{N}(0,\tau_{i}^{-1}), & \tau_{i} & \sim\mathcal{G}(a,b),\,i=1,\ldots,p-1,\label{eq:ard-diff}\\
\beta_{p} & \sim\mathcal{N}(0,\tau_{p}^{-1}), & \tau_{p} & \sim\mathcal{G}(a,b).\label{eq:beta_1}
\end{align}
This approach can be generalized to several dimensions and several
differential operators \cite{chantas2010variational}. 

The problem of combination of these two assumptions is typically solved
by their relative weighting, as done e.g. in the fused lasso. In Bayesian
formulation, this correspond to Gaussian prior on the $\beta$ with
zero mean and covariance matrix in the form of weighed combination
of tridiagonal matrices \cite{Kyung_BFL}. An alternative formulation
is to introduce a correlated prior 
\begin{align}
\beta_{i} & \sim\mathcal{N}(-l_{i}\beta_{i+1},\tau_{i}^{-1}), & \tau_{i} & \sim\mathcal{G}(a,b),i=1,\ldots,p-1,\label{eq:ard-beta-L}
\end{align}
with latent variable $l_{i}$, and $\beta_{p}$ is given by (\ref{eq:beta_1}).
Note that both the sparsity prior (\ref{eq:ard}) and the smoothness
prior (\ref{eq:ard-diff}) are a special case of (\ref{eq:ard-beta-L}),
the former for $l_{i}=0$ and the latter for $l_{i}=-1$. Since $l_{i}$
itself is a regression coefficient, we choose conjugate prior in the
form 
\begin{align}
l_{i} & \sim\mathcal{N}(l_{0},\psi_{i}^{-1}), & \psi_{i} & \sim\mathcal{G}(c,d),\forall i,\label{eq:ard-l}
\end{align}
where $l_{0}$ is a chosen mean (typically between $0$ and $-1$
to favor either sparsity or smoothness) and $\psi_{i}$ is the precision
with Gamma prior. Note that (\ref{eq:ard-l}) is in the form of sparsity
prior to promote minimum differences from the chosen mean $l_{0}$.

The likelihood of the regression model is the conventional 
\begin{align}
y & \sim\mathcal{N}\left(X\beta,\sigma^{-1}I_{n}\right), & \sigma & \sim\mathcal{G}\left(a,b\right),\label{eq:likelihood}
\end{align}
finalizing the full hierarchical model studied here.

Note that the multivariate distribution of vector $\beta$ is then
\begin{align}
\beta|\tau,l & \sim\mathcal{N}\left(\mathbf{0}_{p\times1},\left(L\cdot D\cdot L^{T}\right)^{-1}\right),\label{eq:beta_vec}\\
L & =\begin{bmatrix}1 & 0 & \cdots & 0\\
l_{1} & 1 & 0 & \vdots\\
0 & \ddots & \ddots & 0\\
0 & \ddots & l_{n-1} & 1
\end{bmatrix}
\end{align}
where $L$ is a bidiagonal lower triangular matrix and $D$ is a diagonal
matrix with elements $D_{i,i}=\tau_{i}$. This corresponds to the
tridiagonal covariance matrix of a hierarchical model for fused lasso
model of \cite{Kyung_BFL} with the distinction of different parametrization.
An advantage of the presented form is that both sampling from (\ref{eq:beta_vec})
and its variational inference are trivial. 

\section{Inference of the LS-APC model}

Derivation of the inference algorithm for the model with smoothness
and sparsity prior is relatively simple since the model choices in
the prior are motivated predominantly by conjugacy. The conditional
posteriors are then analytically tractable which allows derivation
of the Gibbs sampling and Variational Bayes. These two methods are
closely related as explained in \cite{VB_Ormerod}. Specifically,
the conditional posteriors for all unknowns are as follows
\begin{align}
\beta|y,\sigma,\tau,l,\psi & \sim\mathcal{N}\left(\mu,\Sigma\right), & \tau|y,\beta,\sigma,l,\psi & \sim\prod_{i=1}^{p}\mathcal{G}\left(\gamma_{i},\delta_{i}\right),\label{eq:gibbs-beta}\\
l|y,\beta,\sigma,\tau,\psi & \sim\prod_{i=1}^{p-1}\mathcal{N}\left(\pi_{i},\rho_{i}^{-1}\right), & \psi|y,\beta,\sigma,\tau,l & \sim\prod_{i=1}^{p-1}\mathcal{G}\left(\lambda_{i},\omega_{i}\right).\nonumber \\
\sigma|y,\beta,\tau,l,\psi & \sim\mathcal{G}\left(\gamma_{\sigma},\delta_{\sigma}\right),\label{eq:gibbs-sigma}
\end{align}
with their shaping parameters 
\begin{align*}
\Sigma & =\left(X^{T}\sigma X+LDL^{T}\right)^{-1}, & \mu & =\Sigma X^{T}\sigma y,\\
\gamma_{\sigma} & =a+\frac{n}{2}, & \delta_{\sigma} & =b+\frac{1}{2}\left(y-X\beta\right)^{T}\left(y-X\beta\right),\\
\gamma_{i} & =a+\frac{1}{2}, & \delta_{i} & =b+\frac{1}{2}\left(\beta_{i}+l_{i}\beta_{i+1}\right)^{2},\\
\pi_{i} & =\frac{\psi_{i}l_{0}-\beta_{i}\beta_{i+1}\tau_{i}}{\psi_{i}+\beta_{i+1}^{2}\tau_{i}}, & \rho_{i} & =\psi_{i}+\beta_{i+1}^{2}\tau_{i},\\
\lambda_{i} & =c+\frac{1}{2}, & \omega_{i} & =d+\frac{1}{2}\left(l_{i}-l_{0}\right)^{2}.
\end{align*}
For sake of simplicity, we define $l_{p}=0$ and $\beta_{p+1}=0$.

\subsection{Gibbs Sampler for LS-APC}

Application of the Gibbs sampler is based on drawing samples from
the conditional distributions. The only sensitive operation is drawing
of samples from $\beta$, where it is necessary to perform Cholesky
decomposition which may be problematic in high dimensions. However,
since both additive terms in $\Sigma$ are in product form, it is
possible to write $\Sigma^{-1}=Q^{T}Q$ with $Q=\left[\sqrt{\sigma}X;\sqrt{D}L^{T}\right]$
and use triangularization procedure such as the QR to obtain triangular
matrix $R$ such that $R^{T}R=Q^{T}Q$. 

The key advantage of the Gibbs sampler is that its convergence to
the true posterior is ensured by the ergodic theorem. This means,
that the only error of the Markov chain can arise from stopping at
finite number of samples. With sufficient number of algorithm iterations,
this inaccuracy can be decreased to a level, where it is of little
importance. This, of course, usually requires great amount of computational
time.

The challenge for inference based on Gibbs sampling is evaluation
of the marginal likelihood for Bayesian model selection (or Bayes
factor). Some methods for Gibbs sampler algorithm are based on direct
approximation of the posterior likelihood $p\left(y|M\right)$, where
$M$ is a categorical variable denoting the index of the evaluated
model from the set of all considered models $M\in\left\{ M_{1},\ldots,M_{m}\right\} $.
Utilization of harmonic mean (\cite{Gibbs-overall}) or importance
sampling technique (\cite{Gibbs-marginal_posterior}) are very common,
because no more distribution sampling is necessary. The disadvantage
of these methods is in numerical evaluation of the probability distribution,
because its values are typically indistinguishable from zero in higher
dimensions. In our experiments, we therefore used a method proposed
in \cite{Gibbs-Chib}, which is based on logarithm form of the Bayes's
equation 
\begin{align}
\text{ln}\left(p\left(y|M\right)\right) & =\text{ln}\left(p\left(y|M,\theta\right)\right)+\text{ln}\left(p\left(\theta\right)\right)-\text{ln}\left(p\left(\theta|y\right)\right),\label{eq:loglik-Gibbs}
\end{align}
where all expressions on the right side can be evaluated for a given
parameter $\theta=\theta^{*}$. This requires additional sampling,
but the logarithm form enables more reliable numerical evaluation.

\subsection{Variational Bayes Method for LS-APC}

The Variational Bayes approximation of the posterior distribution
(also known as mean field approximation) is a less accurate approximation
of true posterior than the Gibbs sampler but usually it is much faster
to evaluate. It is derived by minimization of the Kullback-Leibler
divergence from a chosen approximation (product of conditionally independent
posteriors) to the true posterior. 

Specifically for our model, the approximating distribution of all
unknowns $\theta=\left[\beta,\sigma,\tau,l,\psi\right]$ is chosen
as 
\begin{equation}
p\left(\theta|y,X\right)\approx q\left(\beta,\sigma,\tau,l,\psi\right)=q_{\beta}\left(\beta\right)q_{\sigma}\left(\sigma\right)q_{\tau}\left(\tau\right)q_{l}\left(l\right)q_{\psi}\left(\psi\right).\label{eq:VB-condindep}
\end{equation}
Minimum of the KL divergence
\begin{align*}
KL\left[q\left(\theta\right)||p\left(\theta|y,X\right)\right] & =\int q\left(\theta\right)\text{ln}\left[\frac{q\left(\theta\right)}{p\left(\theta|y,X\right)}\right]d\theta
\end{align*}
is obtained in the general form (\cite{VB_Ormerod,VB_Smidl}) 
\begin{align}
q_{\theta_{k}}\left(\theta_{k}\right) & =\text{exp}\left[\text{E}_{\theta_{j},j\neq k}\text{\,ln}\left(p\left(\theta,y,X\right)\right)\right],\label{eq:VB_q}
\end{align}
where $\text{E}_{\theta_{j},j\neq k}$ denotes the mean value of the
argument over all parameters $\theta_{j}$ except for $\theta_{k}$
and $p\left(\theta,y,X\right)$ is the joint distribution of data
$y$ and all parameters and hyperparameters. The resulting approximate
distribution (\ref{eq:VB_q}) are identical to the conditional posteriors
(\ref{eq:gibbs-beta})\textendash (\ref{eq:gibbs-sigma}), i.e. 
\begin{align}
q_{\beta}(\beta) & =\mathcal{N}\left(\mu,\Sigma\right), & q_{\sigma}(\sigma) & \sim\mathcal{G}\left(\gamma,\delta\right),\nonumber \\
q_{\tau}(\tau) & =\prod_{i=1}^{p}\mathcal{G}\left(\gamma_{i},\delta_{i}\right), & q_{l}(l) & \sim\prod_{i=1}^{p-1}\mathcal{N}\left(\pi_{i},\rho_{i}^{-1}\right), & q_{\psi}(\psi) & =\prod_{i=1}^{p-1}\mathcal{G}\left(\lambda_{i},\omega_{i}\right).\label{eq:VB_qs_SSP}
\end{align}

Parameters off all distributions have the same form as in the case
of the Gibbs sampler, however, with conditioning variables replaced
by their expectations
\begin{equation}
\begin{aligned}\Sigma & =\left(X^{T}\text{E}_{\sigma}\left[\sigma\right]X+\text{E}_{\tau,l}\left[LDL^{T}\right]\right)^{-1}, & \mu & =\Sigma X^{T}\text{E}_{\sigma}\left[\sigma\right]y,\\
\gamma_{\sigma} & =a+\frac{n}{2}, & \delta_{\sigma} & =b+\frac{1}{2}\text{E}_{\beta}\left[\left(y-X\beta\right)^{T}\left(y-X\beta\right)\right],\\
\gamma_{i} & =a+\frac{1}{2}, & \delta_{i} & =b+\frac{1}{2}\text{E}_{\beta,l}\left[\left(\beta_{i}+l_{i}\beta_{i+1}\right)^{2}\right],\\
\pi_{i} & =\frac{l_{0}\text{E}_{\psi}\left[\psi_{i}\right]-\text{E}_{\beta}\left[\beta_{i}\beta_{i+1}\right]\text{E}_{\tau}\left[\tau_{i}\right]}{\text{E}_{\psi}\left[\psi_{i}\right]+\text{E}_{\beta}\left[\beta_{i+1}^{2}\right]\text{E}_{\tau}\left[\tau_{i}\right]}, & \rho_{i} & =\text{E}_{\psi}\left[\psi_{i}\right]+\text{E}_{\beta}\left[\beta_{i+1}^{2}\right]\text{E}_{\tau}\left[\tau_{i}\right],\\
\lambda_{i} & =c+\frac{1}{2}, & \omega_{i} & =d+\frac{1}{2}\text{E}_{l}\left[\left(l_{i}-l_{0}\right)^{2}\right].
\end{aligned}
\label{eq:VB-moments}
\end{equation}
Equations (\ref{eq:VB-moments}) together with equations of the expectations
(such as $\text{E}_{l}\left[l_{i}^{2}\right]=\pi_{i}^{2}+\rho_{i}^{-1}$)
form a set of implicit equations that need to be solved. The standard
approach is based on alternating iterative algorithm which monotonically
converges to a local minimum of the KL divergence. In spite of the
possibility of reaching only local minimum, this simple algorithm
often provide satisfactory results. One of our objectives is to validate
its performance for this particular model. 

\subsubsection{Model Selection}

Model selection in the Variational Bayes approximation is based on
decomposition of the logarithm of marginal likelihood $p\left(y,X\right)$
\begin{align}
\text{ln}\left(p\left(y,X\right)\right) & =\sum_{j}\mathcal{L}_{j}+KL\left[q\left(\theta|M\right)q\left(M\right)||p\left(\theta,M|y,X\right)\right]\label{eq:VB-marg-lik}
\end{align}
where $\mathcal{L}_{j}=\int q\left(Z|M_{j}\right)q\left(M_{j}\right)\text{ln}\left(\frac{p\left(\theta,y,X,M_{j}\right)}{q\left(\theta|M_{j}\right)q\left(M_{j}\right)}\right)$
and $q\left(\theta|M_{j}\right)$ is the approximation (\ref{eq:VB-condindep})
obtained for the model $M_{j}$. Term $q\left(M_{j}\right)$ here
denotes the approximation of the marginal likelihood of $j$th model
and it can be shown (\cite{Bishop}), that the KL divergence is minimized
for choice 
\begin{align}
q\left(M_{j}\right) & =p\left(M_{j}\right)\text{exp}\left[\int q\left(\theta|M_{j}\right)\text{ln}\left(\frac{p\left(\theta,y,X|M_{j}\right)}{q\left(\theta|M_{j}\right)}\right)d\theta\right].\label{eq:VB_qM}
\end{align}
The approximate marginal likelihood () is known to provide lower bound
on the true marginal (\ref{eq:VB-marg-lik}) without any guarantees
of its tightness. Once again, we aim to study its validity in comparison
with the Gibbs sampling approach.

\subsection{Competing Techniques}

For evaluation of the proposed methods, we select the closest competitors
which are the fused lasso and its Bayesian version. 

\subsubsection{Fused Lasso}

The least absolute shrinkage and selection operator (Lasso) proposed
by \cite{LASSO_Tibshirani} is essentially a standard least square
problem with a weighted additional $L_{1}$ penalization of regression
coefficients. The penalized likelihood function is then
\begin{align*}
\mathcal{L}\left(\beta\right) & =\left\Vert y-X\beta\right\Vert _{2}^{2}+\lambda\sum_{j=1}^{p}\left|\beta_{j}\right|,
\end{align*}
where $\lambda$ is a chosen constant. Selection of the optimal values
of parameter $\lambda$ is a complex task. Typically a range of possible
values is evaluated in parallel and their suitability is evaluated
with respect to a selected measure, such as the mean square error
to the ground truth, or cross validation. The additional constraint
ensures the sparse estimation of $\beta$ and is quite useful in $p\gg n$
cases, that is when there is more coefficients than observed data.
Those are the reasons why lasso became so popular in many different
fields and applications \cite{LASSO-ref1,LASSO-ref2,LASSO-ref3,LASSO-ref4}.
Lasso estimates can be interpreted from Bayesian perspective as a
posterior mode estimates when the regression coefficients have independent
Laplace prior. Although lasso is a very popular tool, it doesn't work
with possible correlations between coefficients which results in estimates
that are not smooth.

Fused lasso (FL) is an extension proposed by \cite{FusedLasso}, that
takes into account possible relations between regression coefficients.
This is achieved by adding a new term into the loss function that
penalizes differences between two subsequent coefficients
\begin{align*}
\mathcal{L}\left(\beta\right) & =\left\Vert y-X\beta\right\Vert _{2}^{2}+\lambda_{1}\sum_{j=1}^{p}\left|\beta_{j}\right|+\lambda_{2}\sum_{j=2}^{p}\left|\beta_{j-1}-\beta_{j}\right|,
\end{align*}
where $\lambda_{1}$ and $\lambda_{2}$ are chosen constants that
need to be tuned manually or optimized. This simple enhancement has
a great impact on final estimates of regression coefficients, because,
except for sparsity, it enforces their smoothness.

For optimal choice of Fused lasso parameters we use EMcvfusedlasso
function in HDPenReg R package \cite{HDpenreg} for 5 fold cross validation.

\subsubsection{Bayesian Fused Lasso}

Since lasso and other penalized regression methods based on lasso
are, by its nature, frequentist methods, there was an effort to create
a fully Bayesian approach that would face the smoothness problem in
probabilistic manner. \cite{Kyung_BFL} uses the fact, that lasso
estimates can be interpreted as a posterior mode under Laplace prior
and that Laplace distribution can be expressed as a scale mixture
of Gaussian distributions with independent exponentially distributed
variances. This hierarchical prior model is extended to cover the
Fused lasso and more (eg. Elastic net). The Gibbs sampler for the
Bayesian Fused lasso (BFL) is given in \cite{Kyung_BFL}. However,
Variational inference would be problematic since the expectations
of the conditioning variables are not analytically tractable.

\section{Simulation Studies}

\begin{figure}
\begin{centering}
\includegraphics[width=0.5\linewidth]{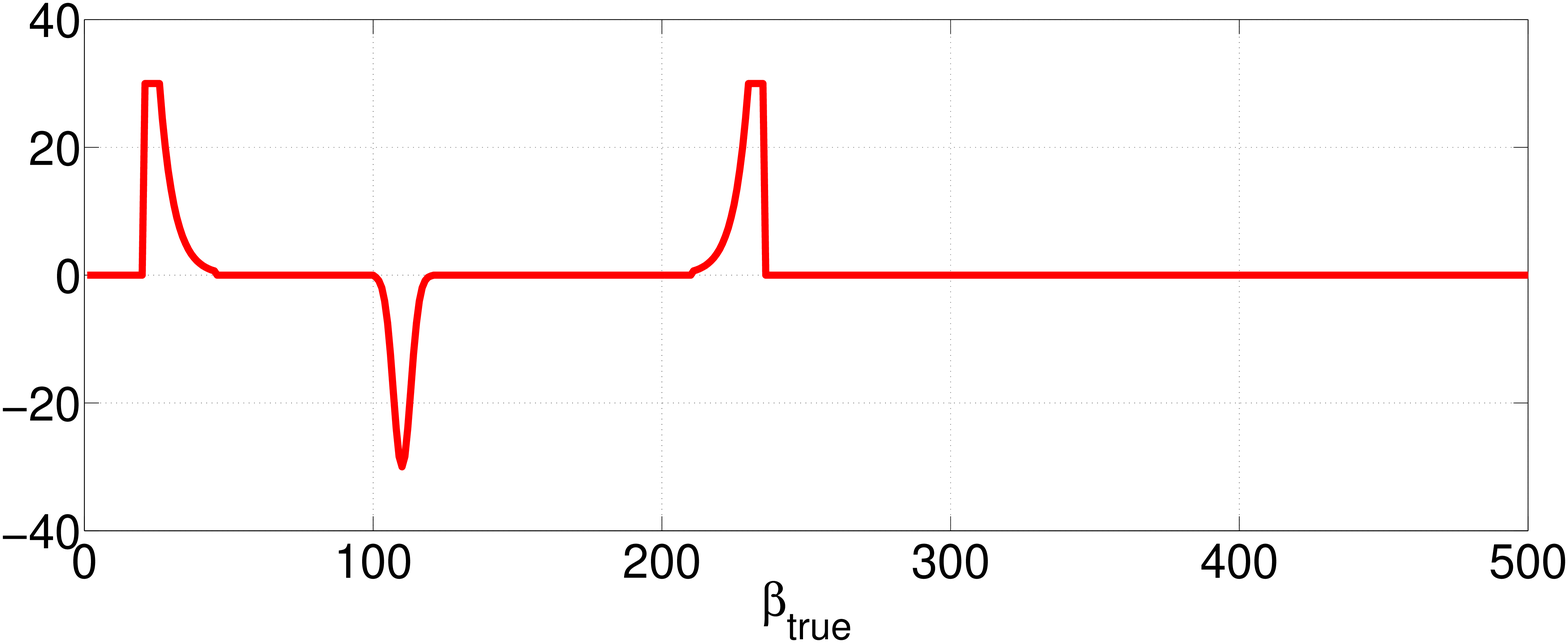}\includegraphics[width=0.5\linewidth]{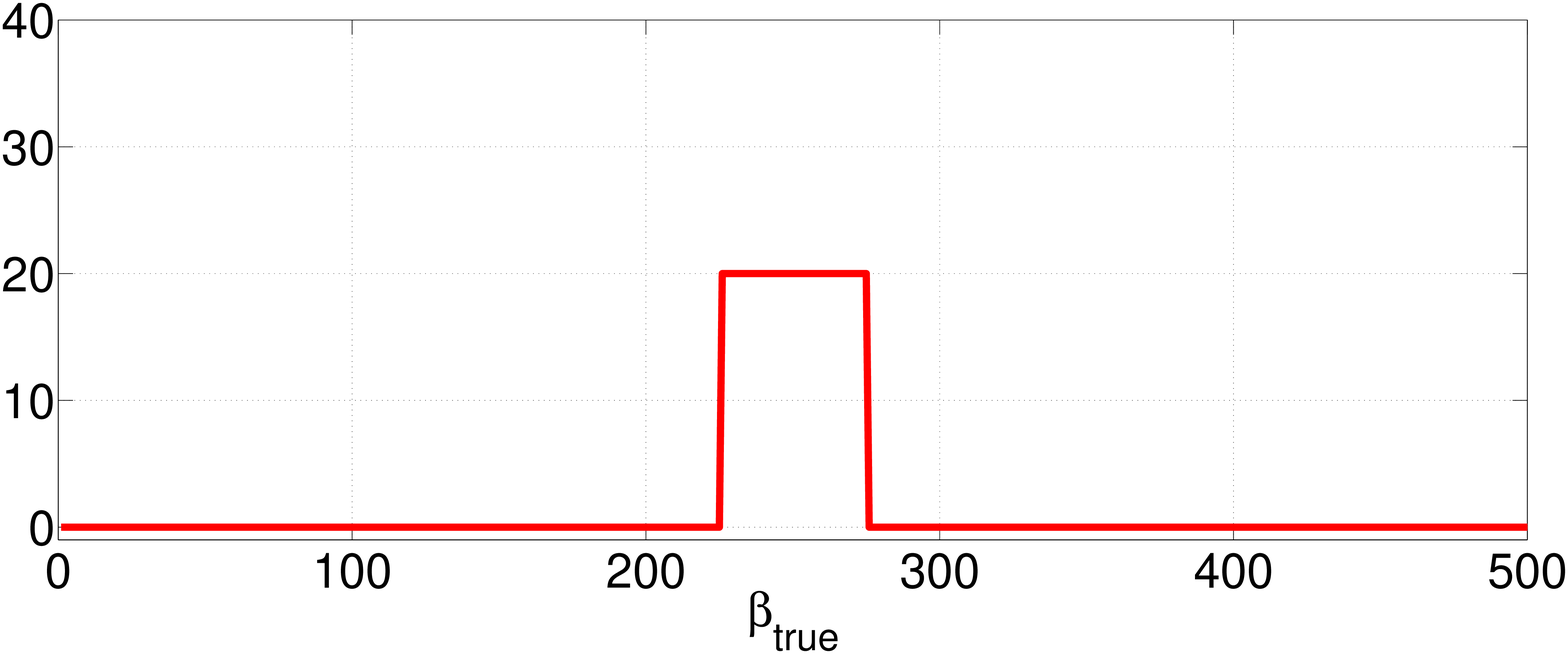}
\par\end{centering}
\caption{\label{fig:beta_true}Simulated ground truth vector of regression
coefficients $\beta$. \textbf{Left}: profile using exponential and
bell shaped curves. \textbf{Right}: piecewise constant profile.}

\end{figure}

We first study the performance of the Fused lasso, Bayesian Fused
lasso, and our LS-APC model (both the Gibbs sampler and Variational
Bayes approximations) on two examples. All of them are based on the
same linear model $y=X\beta+e$, where each entry in matrix $X$ was
simulated from a univariate Gaussian distribution $X_{i,j}\sim\mathcal{N}\left(0,4\right)$
and vector $e$ follows multivariate Gaussian distribution $e\sim\mathcal{N}\left(\boldsymbol{0}_{n\times1},200^{2}I_{n}\right)$.
The ground truth vector of the regression coefficients was chosen
to be both sparse and smooth, its length is $p=500$ and the number
of nonzero coefficients is $70$. These coefficients form three blocks
in the first example, two of these blocks have the shape of exponential
growth and decrease, respectively, and one bell shaped (Gaussian-like)
curve. In the second example, there is only one non-zero constant
block, its length is $50$. The second shape was tested to verify
the tendency of the fused lasso to choose piece-wise constant solutions.
Both chosen shapes of regression coefficients are shown in Figure
\ref{fig:beta_true}.

Performance of the compared methods was tested for different number
of observation. In all cases, the optimal tuning coefficients for
the fused lasso were found using 5 fold cross validation as implemented
in HDPenReg R package, and the number of iterations for both BFL Gibbs
sampler and our LS-APC Gibbs sampler algorithm was 50 000, where the
first 5 000 samples were discarded as a burn in.

Evaluation of all simulation studies is done using absolute error,
i.e. L1 distance of the resulting point estimates from the true parameter:
\begin{equation}
\text{AE}=\sum_{j=1}^{p}\left|\hat{\beta}_{j}-\beta_{j}^{true}\right|\label{eq:AE}
\end{equation}
The point estimate $\hat{\beta}$ is chosen to be the maximum of the
posterior distribution. This is motivated by the results of the Gibbs
samplers, which contain many outliers. The maximum is less sensitive
to the realization of the sampler.

\begin{figure}
\begin{centering}
\includegraphics[width=0.33\linewidth]{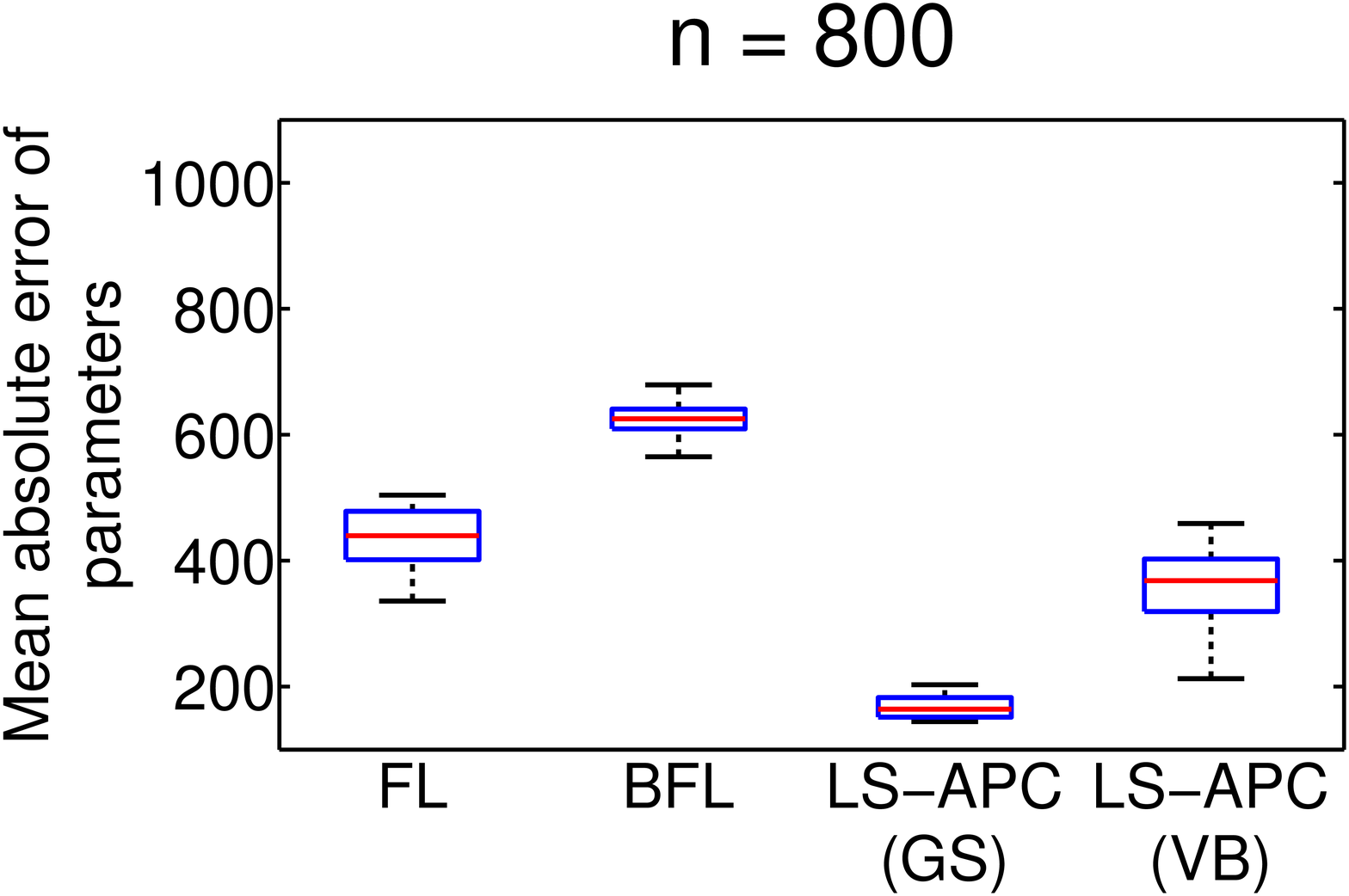}\includegraphics[width=0.33\linewidth]{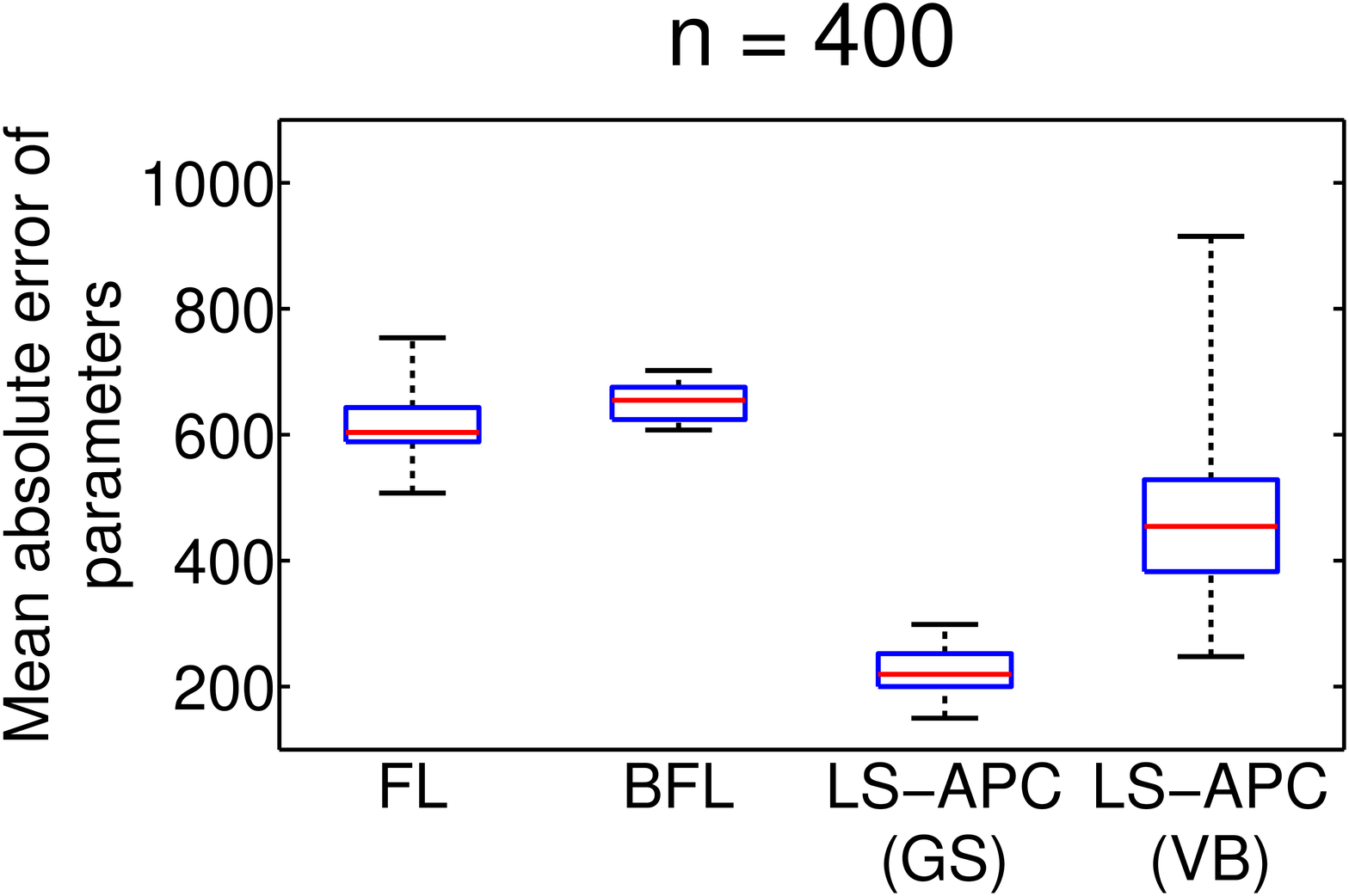}\includegraphics[width=0.33\linewidth]{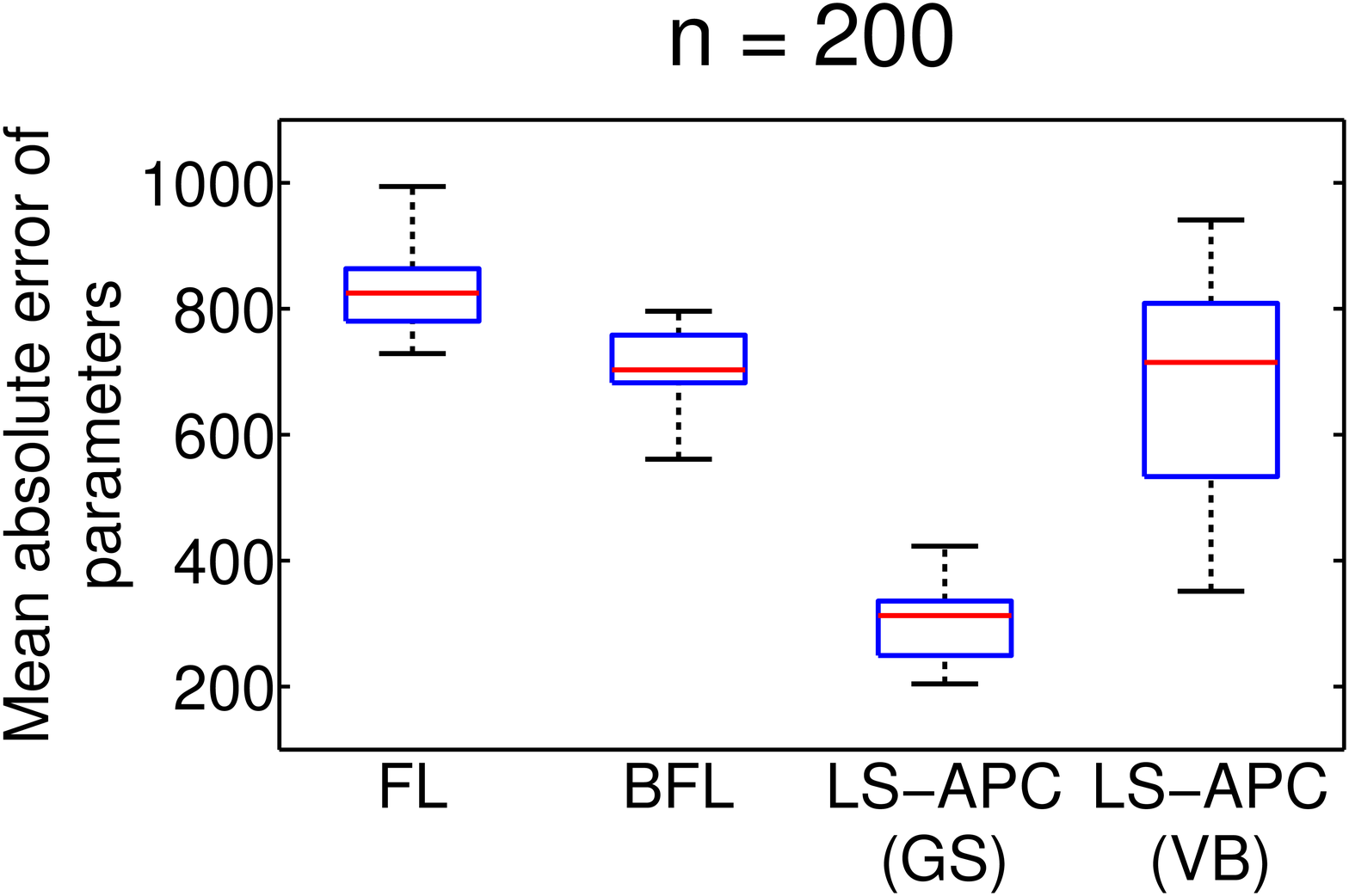}
\par\end{centering}
\caption{\label{fig:ex_curves}Comparison of L1 norm of error of parameter
estimate for different number of observations $n$ of the simulated
model with exponentials and bell curve, Figure \ref{fig:beta_true}
left. Compared methods are: Fused lasso (FL), Bayesian Fused lasso
(BFL), and LS-APC model inferred by the Gibbs sampler (GS) and the
Variational Bayes (VB) respectively.}
\end{figure}

\begin{figure}
\begin{centering}
\includegraphics[width=0.33\linewidth]{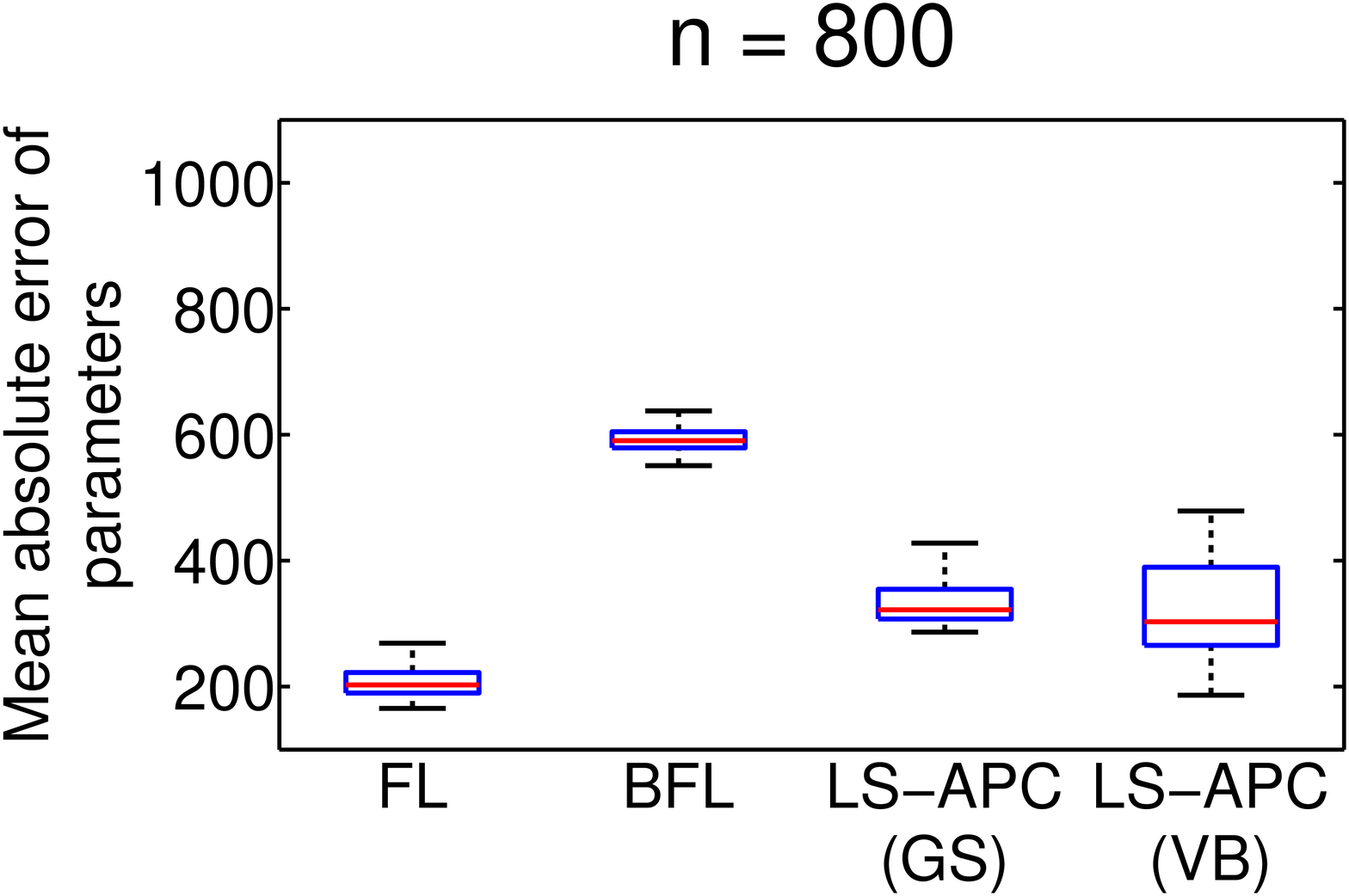}\includegraphics[width=0.33\linewidth]{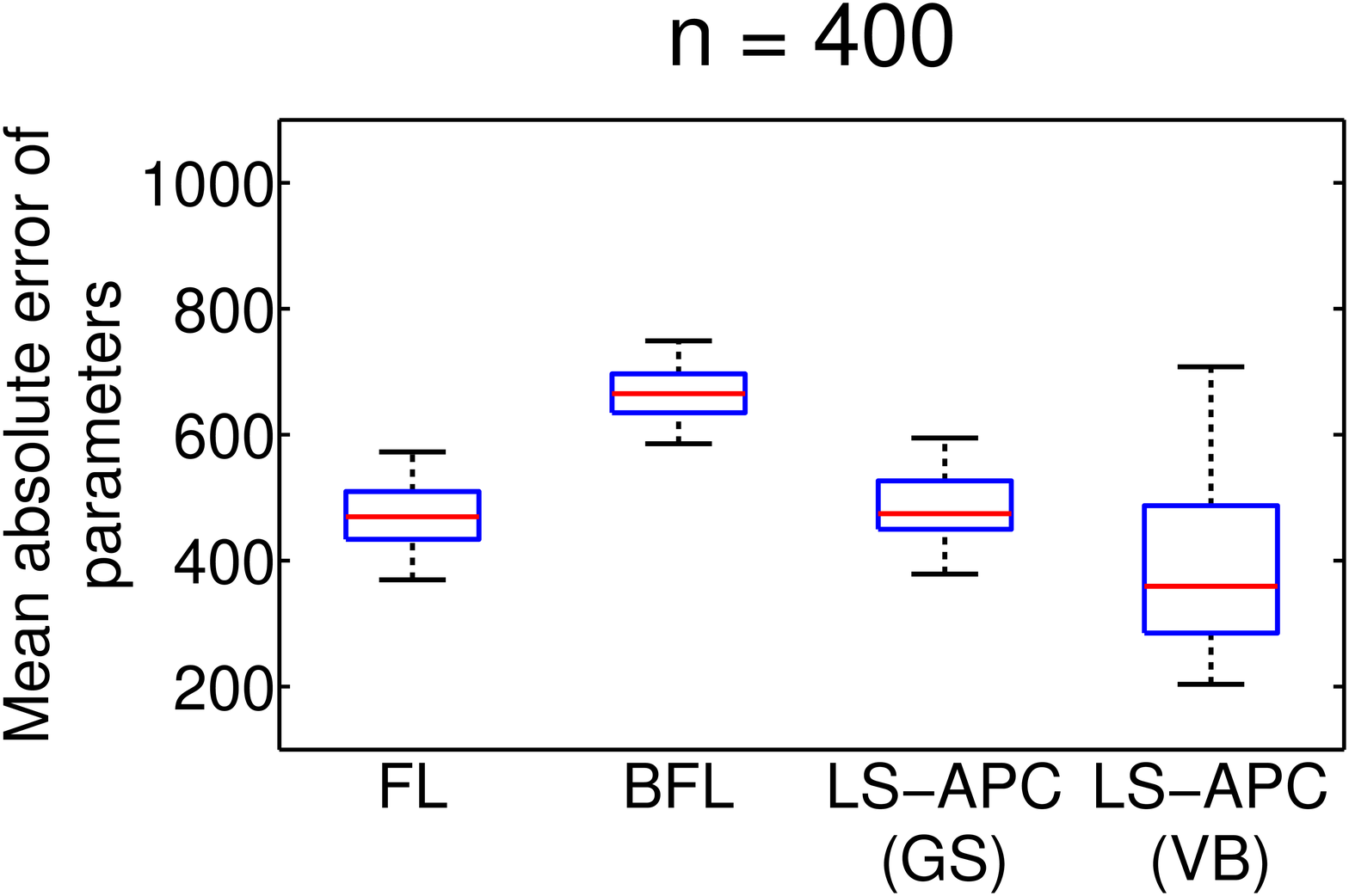}\includegraphics[width=0.33\linewidth]{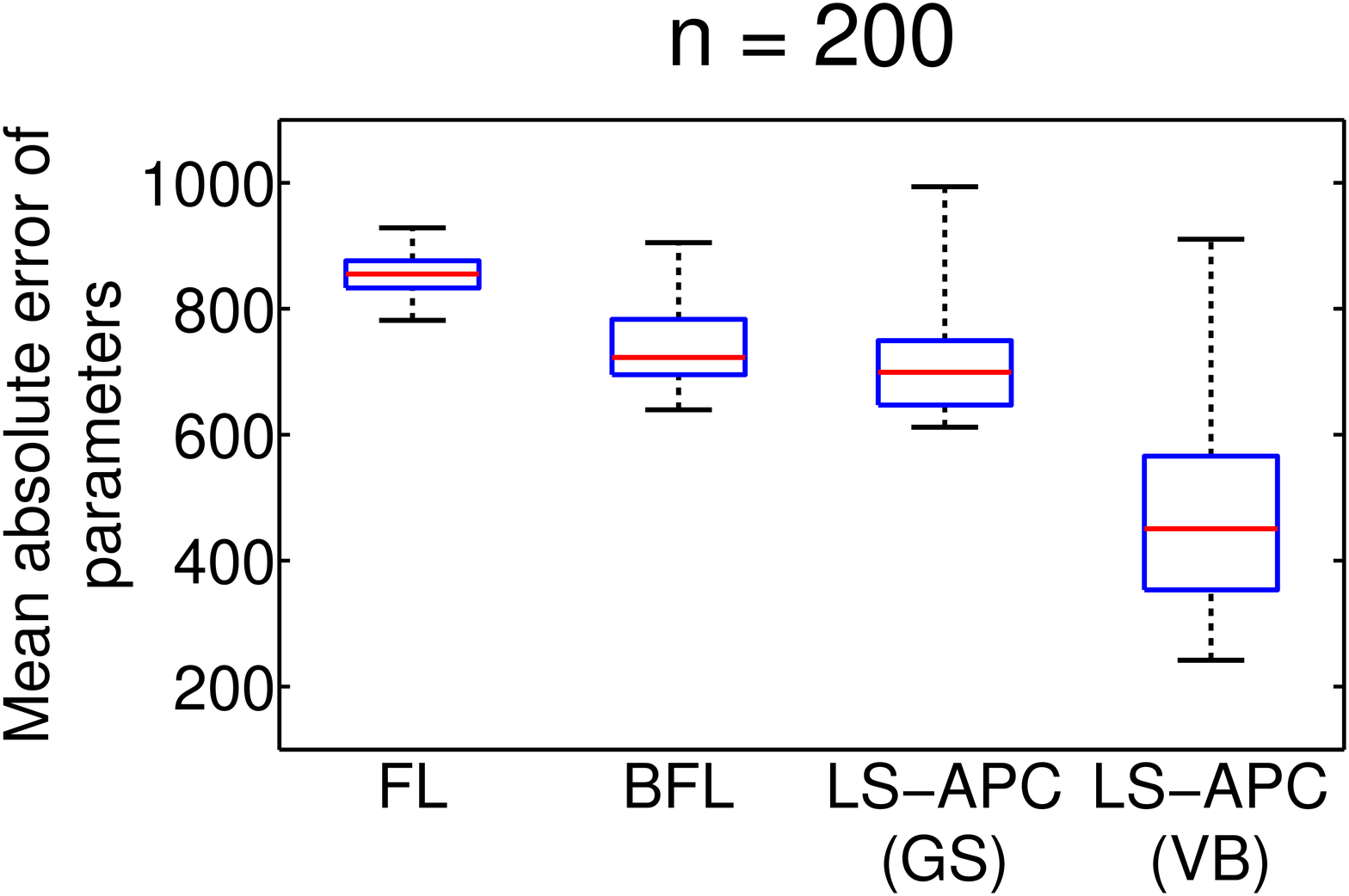}
\par\end{centering}
\caption{\label{fig:ex_constant}Comparison of L1 norm of error of parameter
estimate for different number of observations of the simulated model
with piecewise constant, Figure \ref{fig:beta_true} right. Compared
methods are: Fused lasso (FL), Bayesian Fused lasso (BFL), and LS-APC
model inferred by the Gibbs sampler (GS) and the Variational Bayes
(VB) respectively.}
\end{figure}

Results of Monte Carlo study of the tested methods for 20 realizations
of the linear model with first shape of the parameter (exponentials
and bell curve) is displayed in Figure \ref{fig:ex_curves} for $n\in\left[800,400,200\right]$
observations. Results of the same experiment for the second shape
of the parameter (piecewise constant) are displayed in Figure \ref{fig:ex_constant}.
Indeed, with sufficient amount of observation, the Fused Lasso tuned
by cross-validation outperformed all competing method for piece-wise
constant parameter. However, with lower number observations and for
more complex parameter shape, the Bayesian versions are outperforming
it. The LS-APC variants systematically outperforming the BFL on all
variants. As a mean field approximation, the VB methods is most sensitive
to realization of the measurement error. However, its execution time
is a fraction of all competing methods, therefore it may be a viable
candidate for processing of large data. Note that even its worst performance
is often comparable to that of the FL or BFL.

\section{The European Tracer Experiment Data}

The motivation of our research is inversion modeling of atmospheric
releases of aerosols \cite{stohl2012xenon}, where the LS-APC model
was successfully applied by \cite{Tichy2016gmd}. One of the best
data sets in this field is the ETEX data. The European tracer experiment
(ETEX) were two long time prepared experiments that took place in
autumn 1994 in the north-western part of France. During both of them,
340 kg of perfluoromethylcyclohexane, breathable, well-traceable chemical,
was released and its spread was tracked via a grid of 168 ground stations
located in 17 European countries. The furthest station was more than
2000 km away from the source. An airborne measurement support was
provided by 3 airplanes, but we did not include these observations. 

During the first experiment, from which we have the data available,
340 kg of perflurocarbon was released in period of 12 hours, starting
on 23 October, 16:00 UTC and ending at 04:00 UTC. At each ground station,
automated sequential air samplers operated sampling every 3 hours
for a total period of 72 hours. These samples are arranged in vector
$y$ of length $3102$. We consider the task of estimation of the
temporal profile of the source activity from the known location. The
common methodology is based on running a simulation model with unit
release for each hour in the potential release time window. Since
the chemical is not reactive, the measured concentration is a superposition
of contributions from each hour of the release weighted by the released
amount of the tracer in that hour. The observations are thus explained
by a linear model with regression coefficients given by numerical
simulation model. 

\subsection{Construction of the Regressor Matrix}

We have used version 8.1 of the Lagrangian particle dispersion model
FLEXPART \cite{stohl1998validation} for construction of the matrix
of regressors $X$. This matrix is known as the source-receptor sensitivity
matrix in the atmospheric science. We used the same setup as \cite{Tichy2016gmd,martinez2014robust}
simulating a series of 1hour releases for the period of 5 days containing
the true release. The length of the unknown release profile $\beta$
is thus 120. The model was driven by 40-year re-analysis meteorelogical
data (ERA-40) using model time step in the boundary layer limited
to less than 20\% of the Lagrangian timescale and a maximum value
of 300 s. The simulation was evaluated for all sensors in the observation
network. 

\begin{figure}
\begin{centering}
\includegraphics[width=0.7\linewidth]{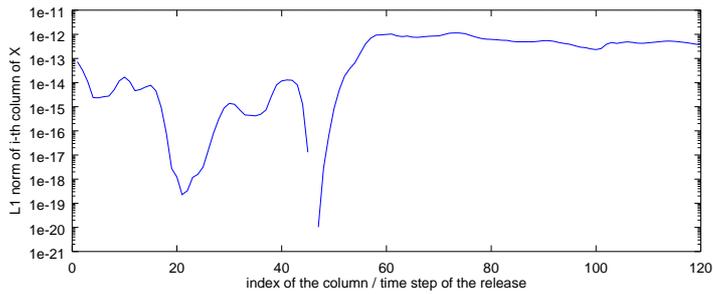}
\par\end{centering}
\caption{\label{fig:Visualization-of-matrix}Visualization of matrix $X$ via
logarithm of the L1 norm of its columns.}

\end{figure}

The regressor matrix $X$ is poorly conditioned as demonstrated in
Figure \ref{fig:Visualization-of-matrix} via L1 norm of its columns.
Note that in time step 46, the sum of the regressors is exactly zero.
Therefore, any value of $x_{46}$ will contribute zero to the measurements.
Sensitivity of the measurements to parameter values with indeces lower
than 54 is also very low. Therefore, we will study behavior of the
methods also for 66 hours with time index $54$ to $120$, this reduced
dataset will be denoted as ETEX 66.

\subsection{Covariance Matrix of the Observations}

Since the matrix $X$ is computed from estimated meteorological data,
it may be subject to a systematic error. Thus, the observation error
may not be independent. For example, if the estimated wind speed is
different from the true wind speed, the model residues would be correlated
in time. Since we can not estimate the full matrix of observations,
we define a finite set of covariance matrices of the observation noise
and use them as models in the model selection procedure.

Specifically, we consider the noise to be Gaussian $e\sim\mathcal{N}\left(\boldsymbol{0}_{n\times1},\sigma^{-1}B^{-1}\right)$,
with 
\begin{equation}
B_{i,j}=\begin{cases}
1 & \text{if }i=j,\\
\xi & \text{site}\left(i\right)=\text{site}\left(j\right)\\
0 & \text{otherwise}.
\end{cases}\text{ and }|\text{time}\left(i\right)-\text{time}\left(j\right)|=3\text{hours},\label{eq:B}
\end{equation}
Hyperparameter $\xi$ is used to model correlation between every two
measurements of the same ground station that are taken in the subsequent
time samples (i.e. 3 hours). The value of $\xi$ also influences positive
definiteness of matrix $B^{-1}$ restricting its possible values.
We have selected a finite grid of $\xi$ and for each value we perform
data transformation
\begin{align}
\tilde{y}_{k} & =\text{chol}\left(B\left(\xi_{k}\right)\right)^{-1}y, & \tilde{X}_{k} & =\text{chol}\left(B\left(\xi_{k}\right)\right)^{-1}X,\,\,k=1,\ldots,m.\label{eq:yX_trans}
\end{align}
Under the assumption that the model is correct, the transformed data
satisfy model (\ref{eq:likelihood}). Note that the conventional uncorrelated
noise is a special case of (\ref{eq:B}) for $\xi=0$.

Remark: We have defined a much larger set of covariance matrices considering
correlation of locations (which could be caused e.g. by error in the
wind direction), or weighting by distance of the receptors. None of
these models was found to be significant. Therefore, we report only
results of the time-correlated observation covariance. 

\subsection{Positivity Enforcement}

Another important feature of the ETEX data is that the released amount
of the tracer can not be negative. Thus the support of the prior on
$\beta$ is constrained to positive real numbers. The prior of $\beta$
is then a truncated multivariate Gaussian distribution. This influences
evaluation of the posterior in both estimation techniques. Specifically,
in the Variational Bayes approach, the shaping distributions are identical
to those with unconstrained version, however, the moments are evaluated
with respect to the truncated region \cite{Tichy2016gmd}.

The Gibbs sampler requires to sample from the multivariate truncated
Gaussian distribution. We have used algorithm of \cite{GS_trunc_sampling}
for this purpose.

\subsection{Model Selection}

Different transformations of the data (\ref{eq:yX_trans}) allow us
to compare performance of both tested inference methods for model
selection of the LS-APC model: the Variational Bayes and the Gibbs
sampler. Since the marginal likelihood values were extremely low,
we report the logarithm of the marginal likelihood relative to its
maximum value for each method in Table \ref{tab:Comparison of log(p(Y))}
and Figure \ref{fig:Comparison of log(p(Y))}. For comparison, we
also evaluated the Gibbs sampling for two choices of $\theta^{*}$:
the maximum aposteriori estimate and the median estimate. Note that
all inference methods decisively select the model with $\xi=0.45$
as the most likely model. 

The correspondence of the model likelihood from the Variational Bayes
method with that of the Gibbs sampler is remarkably good. The Variational
Bayes tends to underestimate the variance of the estimate (\cite{bishop2006pattern}),
which may be the reason why the marginal log-likelihood is underestimated
for the less likely models. 
\begin{center}
\begin{table}
\begin{centering}
{\small{}}%
\begin{tabular}{ccccccccccc}
\toprule 
{\small{}$\xi$} &
{\small{}-0.2} &
{\small{}-0.1} &
{\small{}0} &
{\small{}0.1} &
{\small{}0.2} &
{\small{}0.3} &
{\small{}0.35} &
{\small{}0.4} &
{\small{}0.45} &
{\small{}0.5}\tabularnewline
\midrule
{\small{}GS med.} &
{\small{}-421} &
{\small{}-459} &
{\small{}-469} &
{\small{}-387} &
{\small{}-386} &
{\small{}-323} &
{\small{}-275} &
{\small{}-174} &
{\small{}0} &
{\small{}-137}\tabularnewline
\midrule 
{\small{}GS MAP} &
{\small{}-355} &
{\small{}-347} &
{\small{}-377} &
{\small{}-396} &
{\small{}-389} &
{\small{}-331} &
{\small{}-256} &
{\small{}-128} &
{\small{}0} &
{\small{}-62}\tabularnewline
\midrule 
{\small{}VB} &
{\small{}-673} &
{\small{}-651} &
{\small{}-618} &
{\small{}-570} &
{\small{}-497} &
{\small{}-378} &
{\small{}-265} &
{\small{}-146} &
{\small{}0} &
{\small{}-92}\tabularnewline
\bottomrule
\end{tabular}
\par\end{centering}{\small \par}
\caption{\label{tab:Comparison of log(p(Y))}Comparison of marginal log-likelihood
of the LS-APC model for varying value of $\xi$.}
\end{table}
\begin{figure}
\begin{centering}
\includegraphics[scale=0.25]{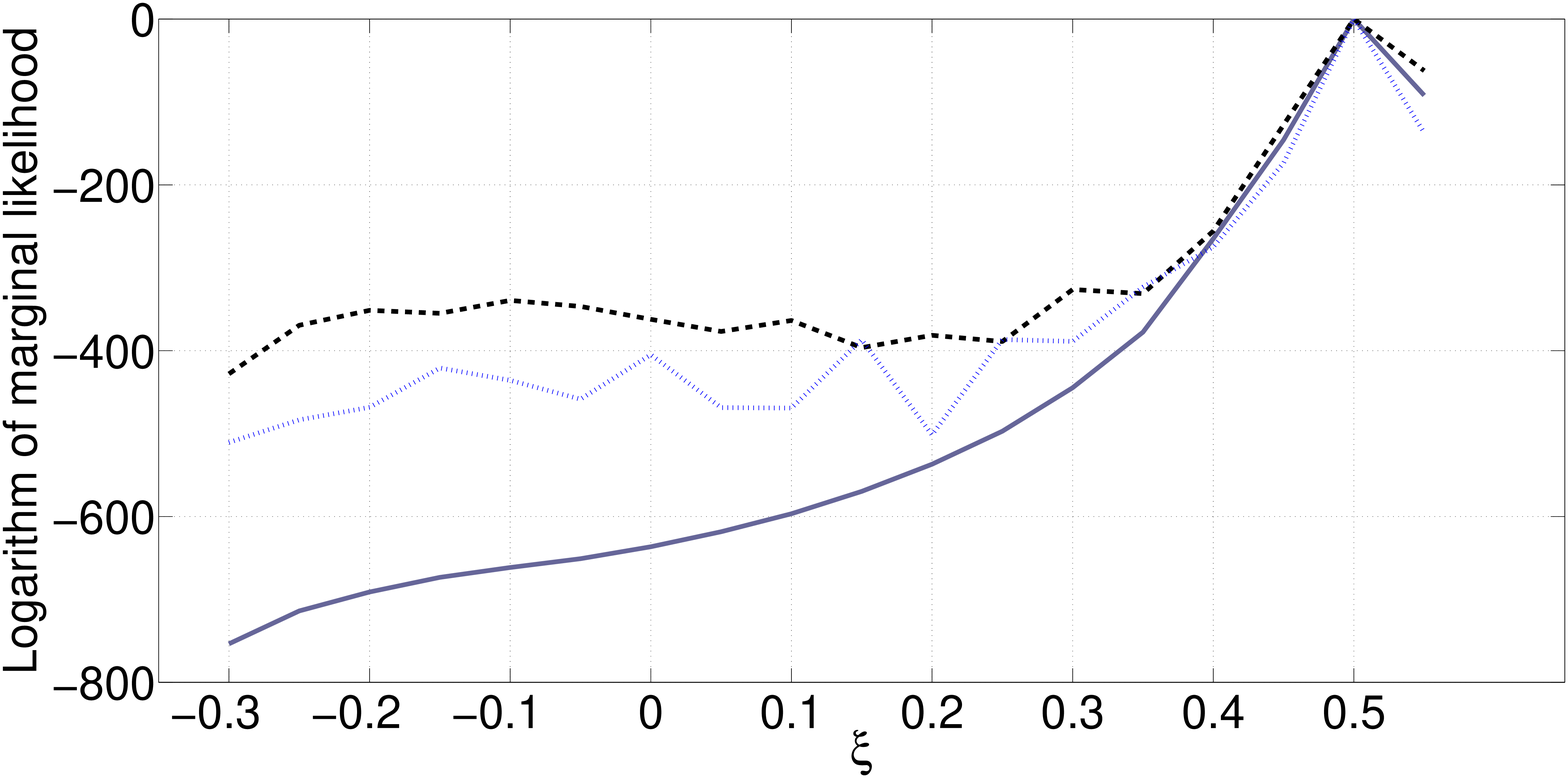}
\par\end{centering}
\caption{\label{fig:Comparison of log(p(Y))}Comparison of marginal log-likelihood
of the LS-APC model for varying value of $\xi$. Dashed line denotes
values obtained by Gibbs sampler with maximum posterior values as
$\theta^{*}$ in (\ref{eq:loglik-Gibbs}) and dotted line denotes
the same estimate using median as $\theta^{*}$. Full line denotes
the results of the Variational Bayes method.}

\end{figure}
\par\end{center}

\subsection{Estimated Source Term Profiles}

In this Section, we provide results of estimation of the release profile
for the ETEX data using all tested algorithms for error covariance
model (\ref{eq:B}) with $\xi=0.45$. 

\begin{figure}
\noindent \begin{centering}
\begin{tabular}{>{\centering}p{0.48\linewidth}>{\centering}p{0.48\linewidth}}
Fused lasso &
Bayesian Fused lasso\tabularnewline
\includegraphics[width=1\linewidth]{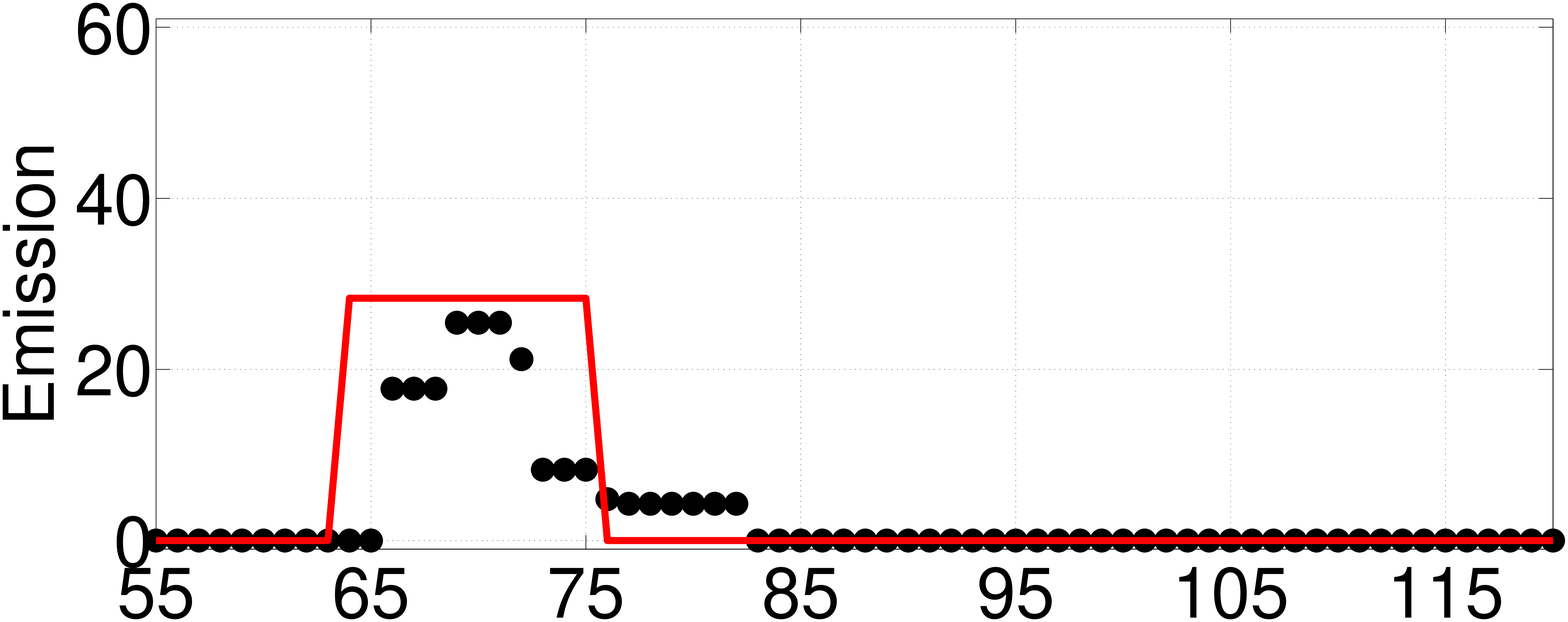}\vspace{-2cm}
 &
\includegraphics[width=1\linewidth]{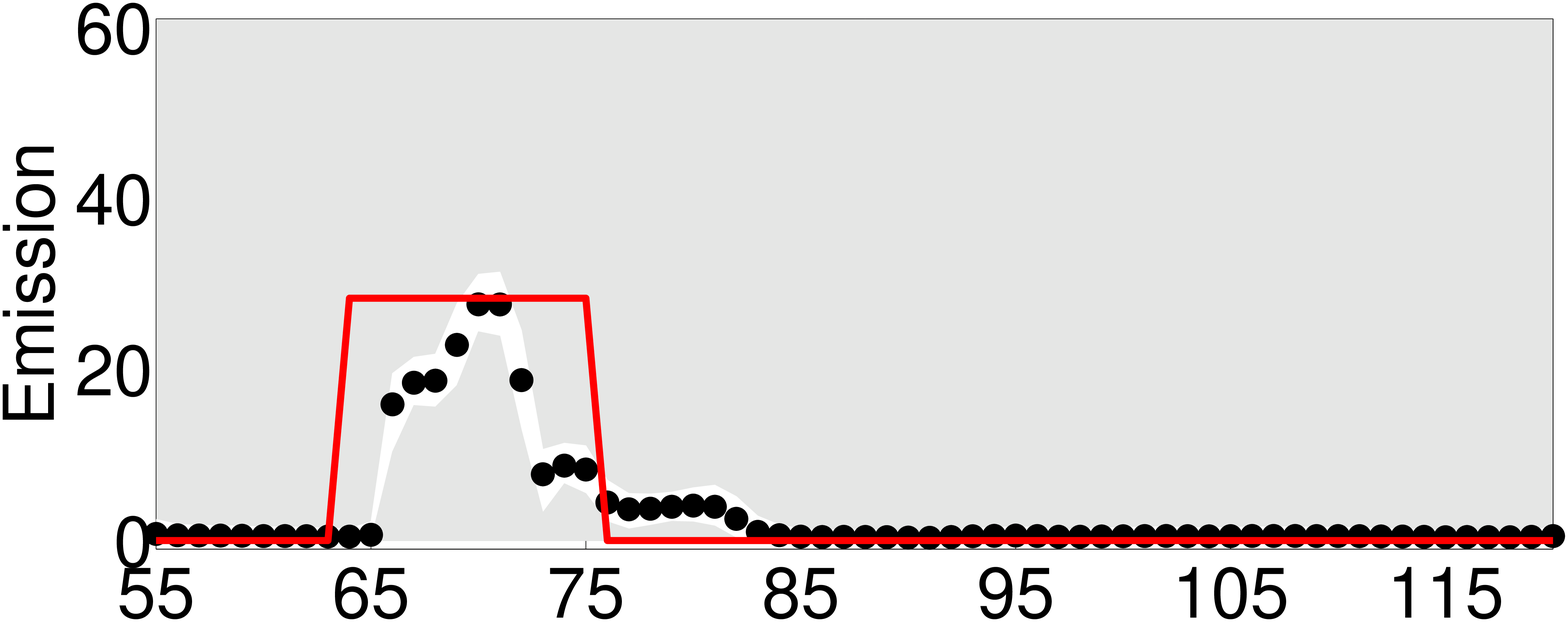}\vspace{-2cm}
\tabularnewline
\hline 
\vspace{-0.15cm}
 &
\vspace{-0.15cm}
\tabularnewline
LS-APC (GS)  &
LS-APC (VB) \tabularnewline
\includegraphics[width=1\linewidth]{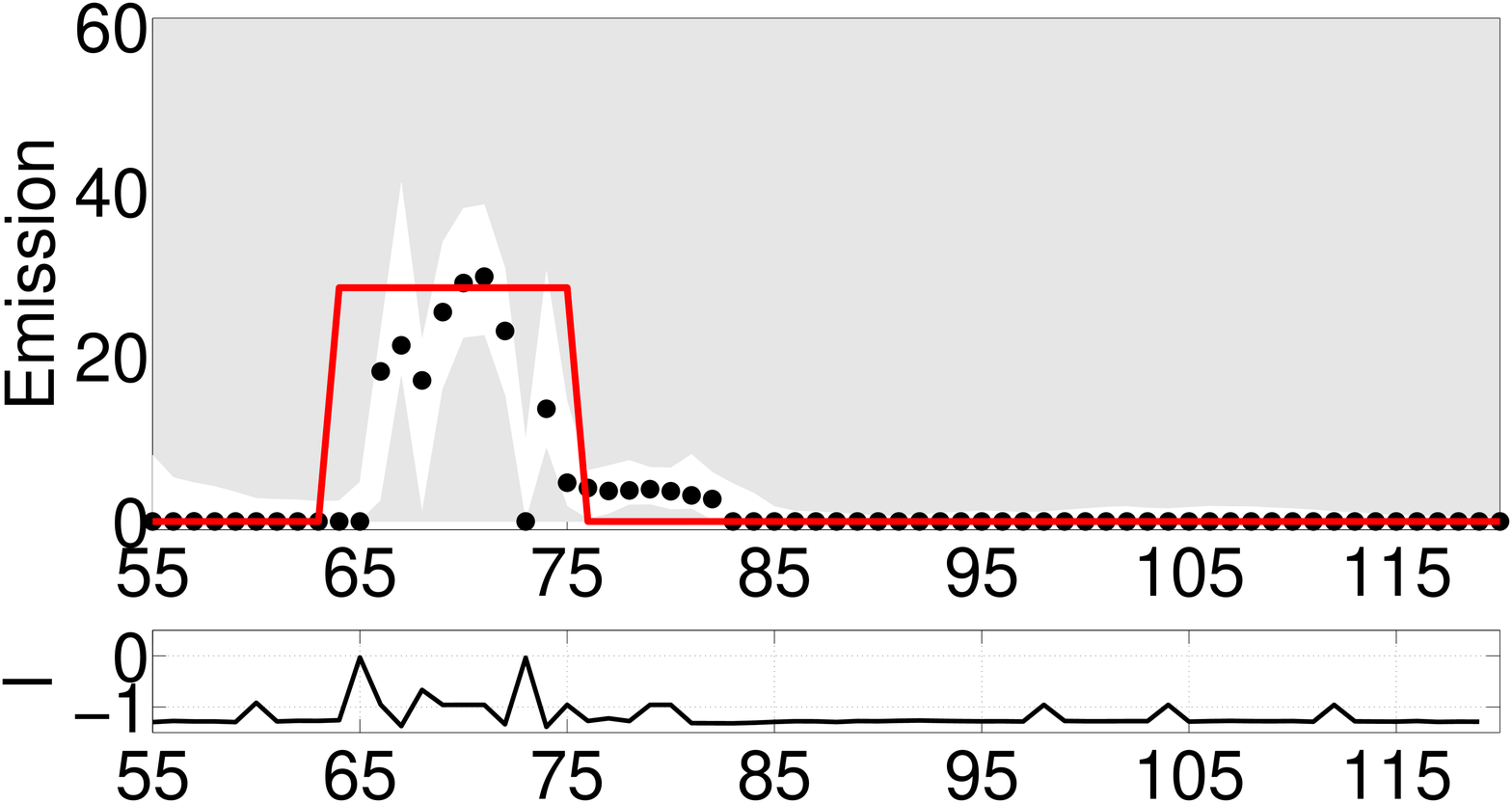} &
\includegraphics[width=1\linewidth]{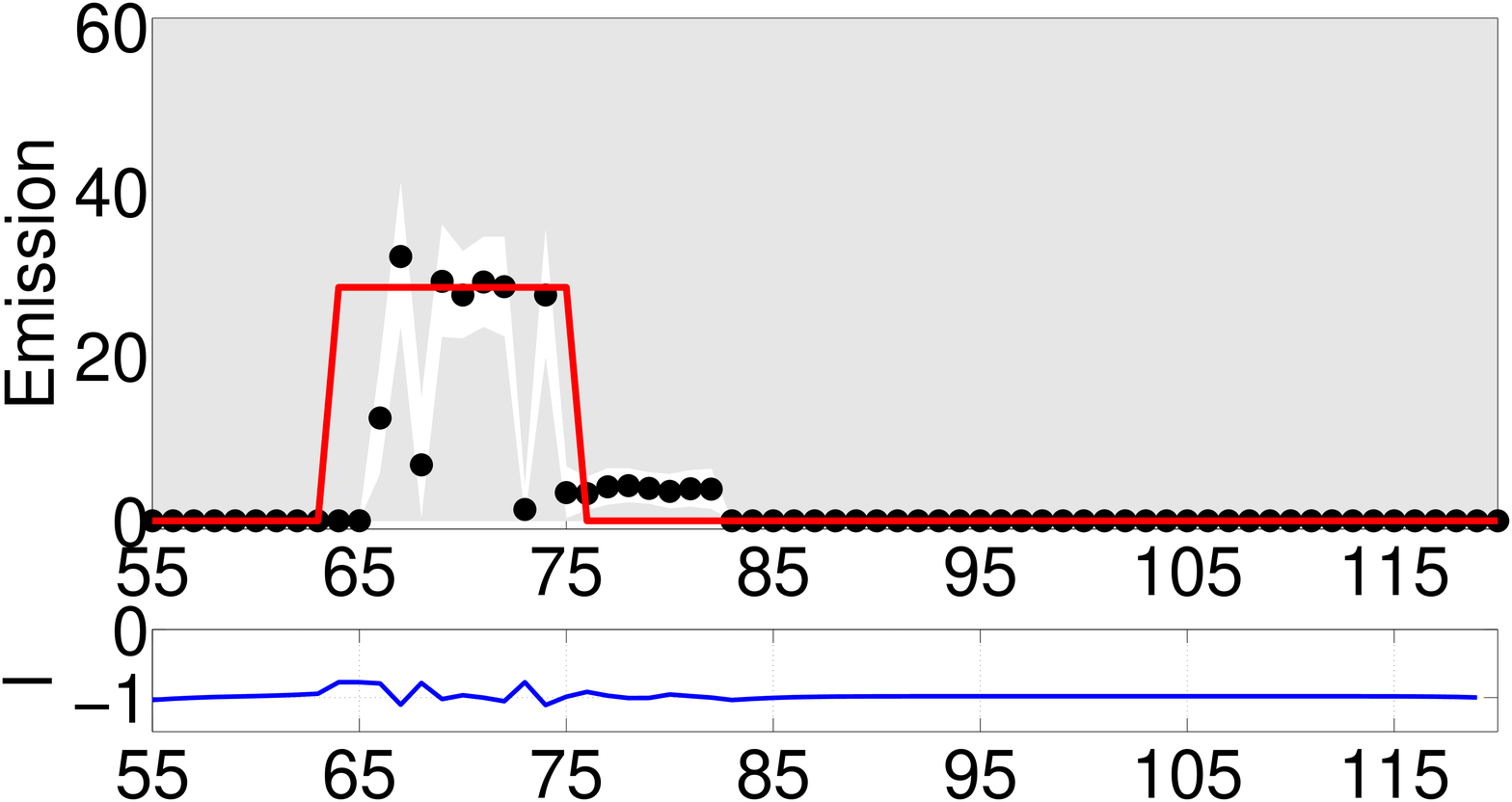}\tabularnewline
\hline 
\vspace{-0.15cm}
 &
\vspace{-0.15cm}
\tabularnewline
LS-APC (GS) without smoothness  &
LS-APC (VB) without smoothness\tabularnewline
\includegraphics[width=1\linewidth]{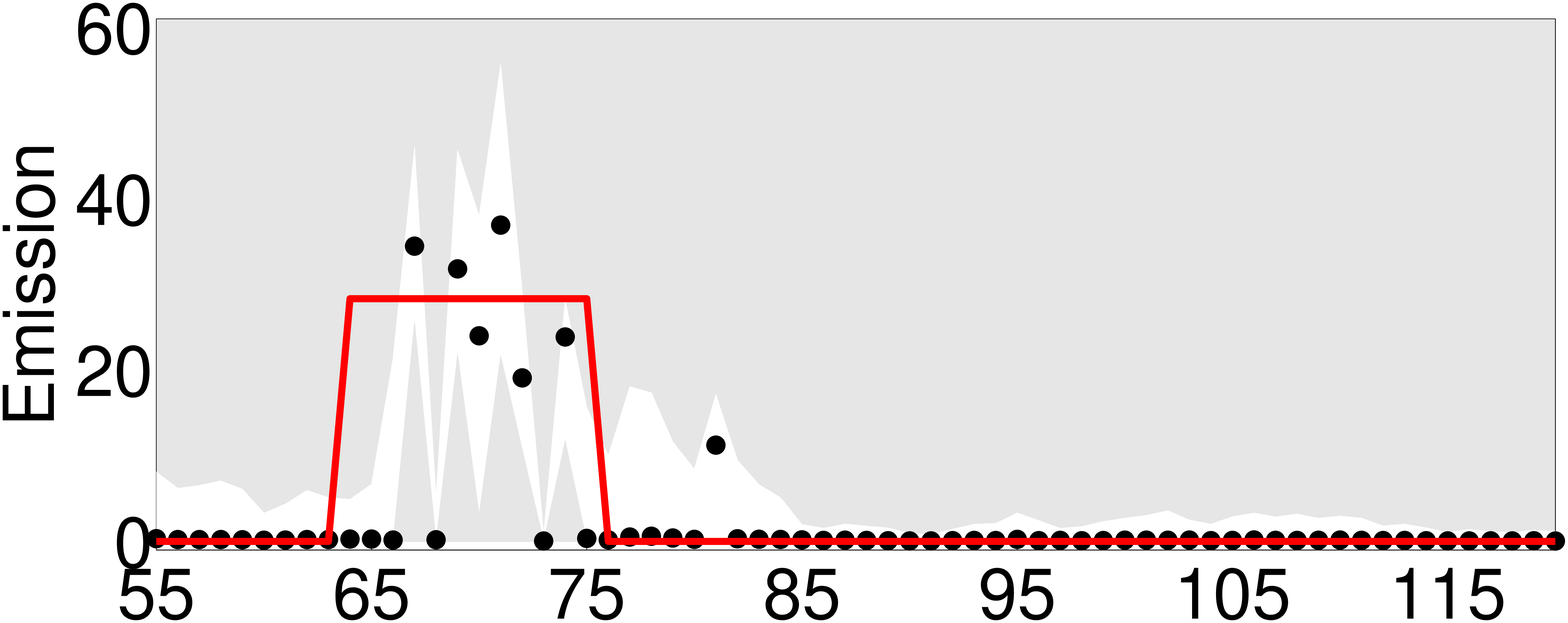}\vspace{-0.3cm}
 &
\includegraphics[width=1\linewidth]{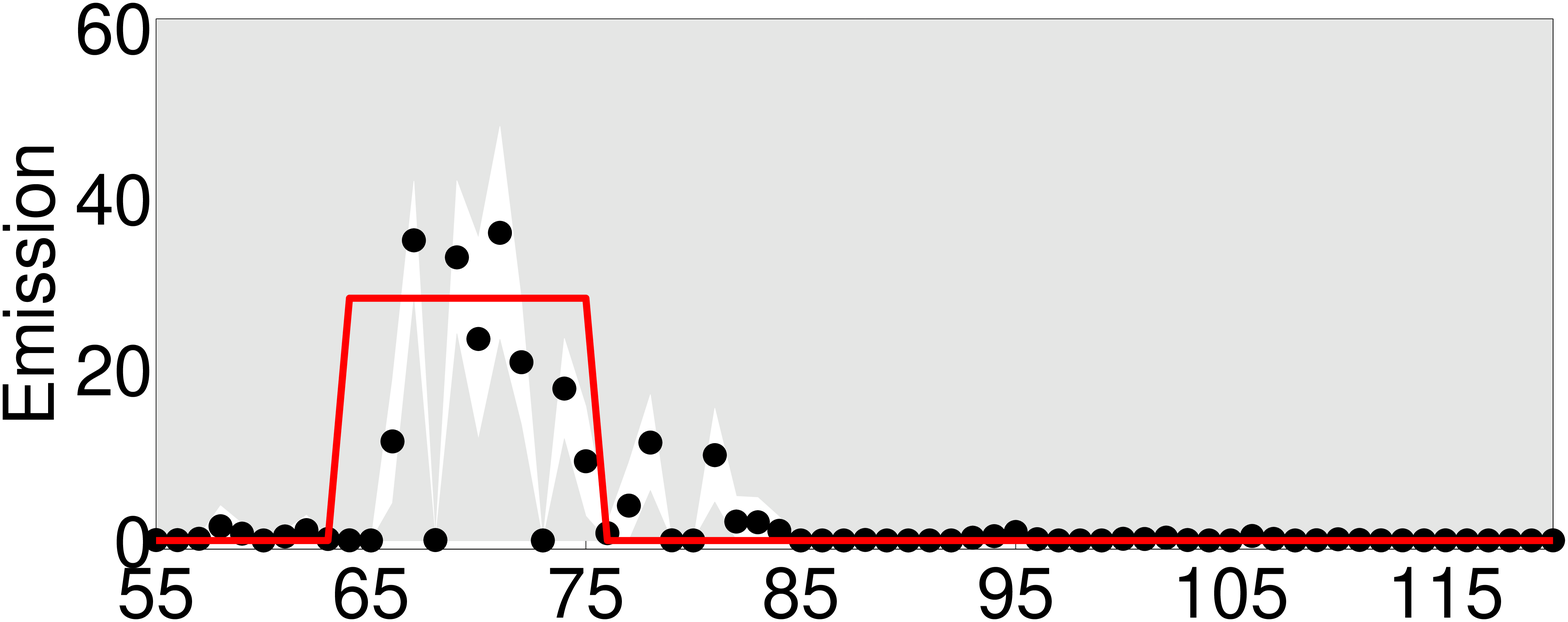}\vspace{-0.3cm}
\tabularnewline
\end{tabular}
\par\end{centering}
\caption{\label{fig:ETEX66}Posterior distribution of the release profile for
the ETEX 66 data set via its maximum and 95\% quantile. In all pictures,
the red line marks the ground truth for $\beta$, black circles are
the MAP estimate of $\beta$ and white area is the 95\% quantile.
The results of LS-APC method are accompanied by estimate of the correlation
parameter $l$ in the lower plot.}
\end{figure}

\begin{figure}
\begin{centering}
\begin{tabular}{>{\centering}p{0.5\linewidth}>{\centering}p{0.5\linewidth}}
Fused lasso &
Bayesian Fused lasso\tabularnewline
\includegraphics[width=1\linewidth]{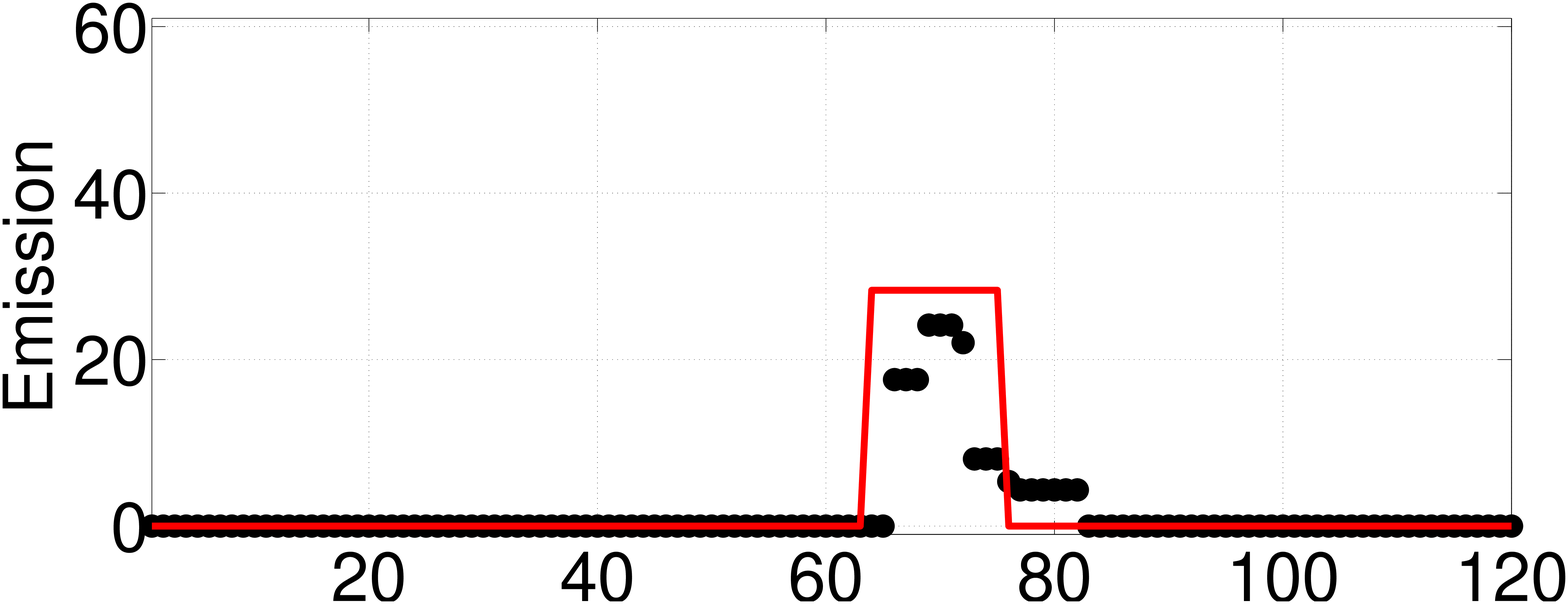}\vspace{-2cm}
 &
\includegraphics[width=1\linewidth]{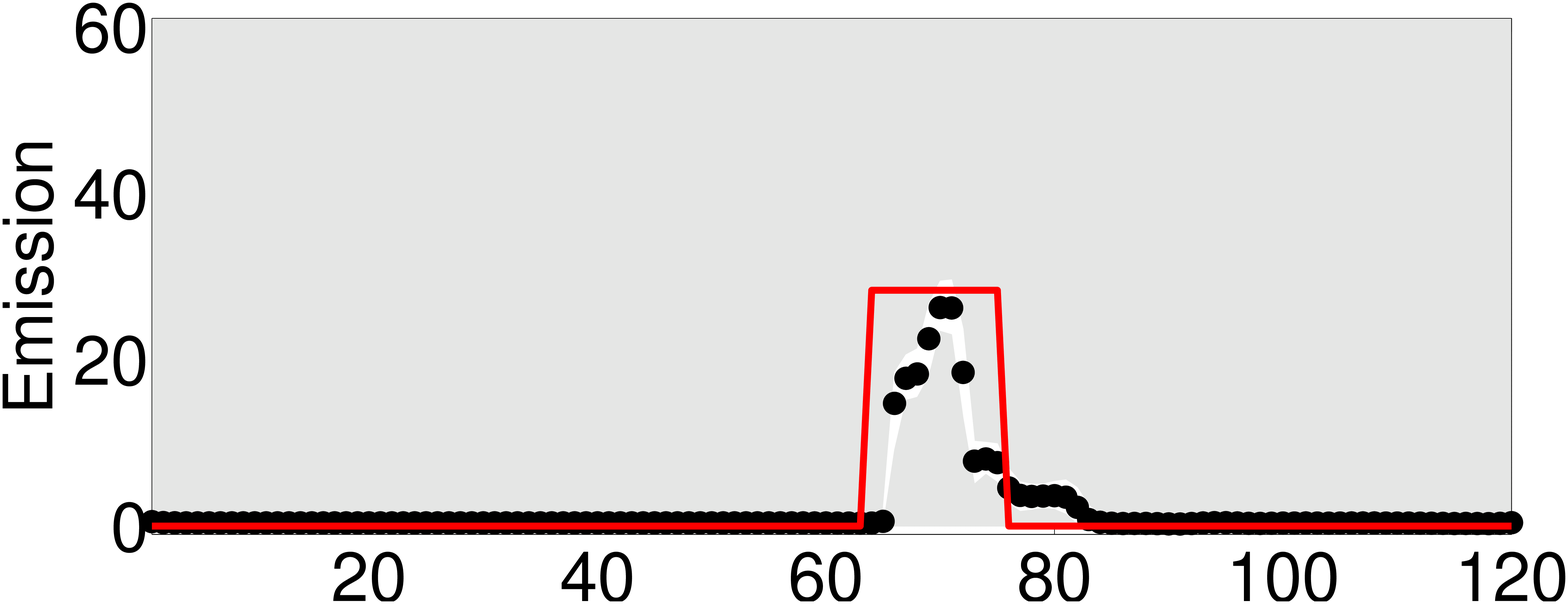}\vspace{-2cm}
\tabularnewline
\hline 
\vspace{-0.15cm}
 &
\vspace{-0.15cm}
\tabularnewline
LS-APC (GS) &
LS-APC (VB)\tabularnewline
\includegraphics[width=1\linewidth]{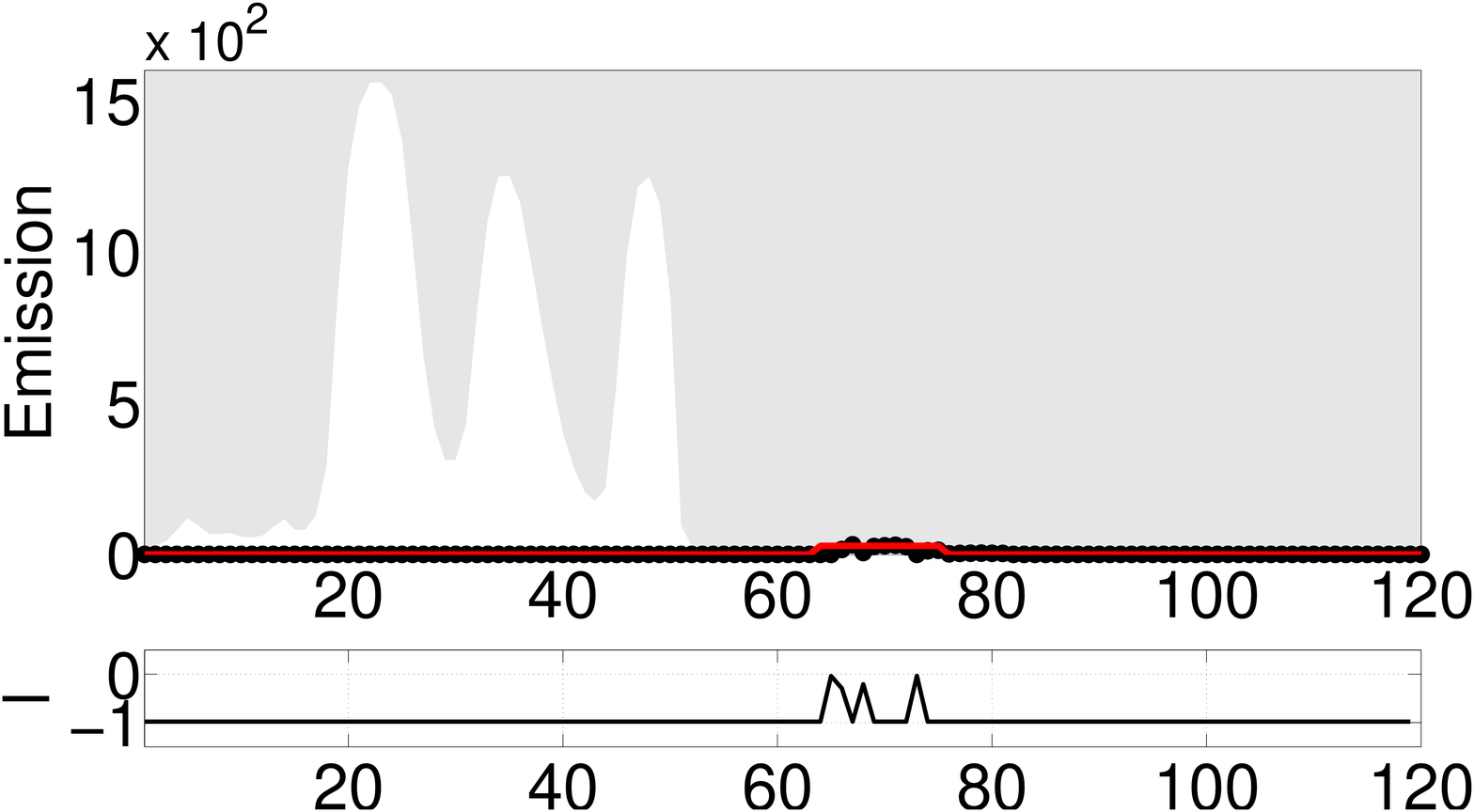} &
\includegraphics[width=1\linewidth]{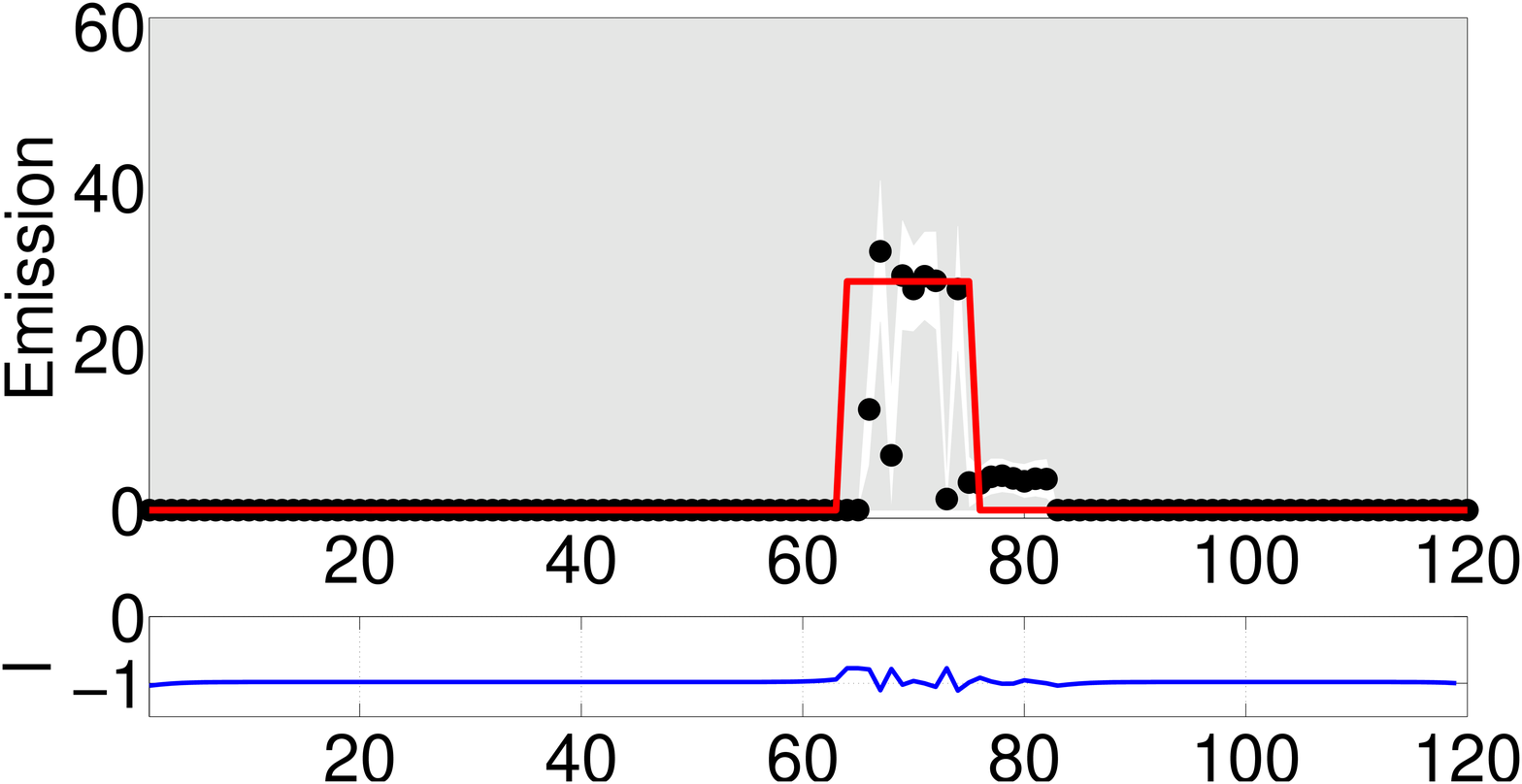}\tabularnewline
\hline 
\vspace{-0.15cm}
 &
\vspace{-0.15cm}
\tabularnewline
LS-APC (GS) without smoothness $l=0$ &
LS-APC (VB) without smoothness $l=0$\tabularnewline
\includegraphics[width=1\linewidth]{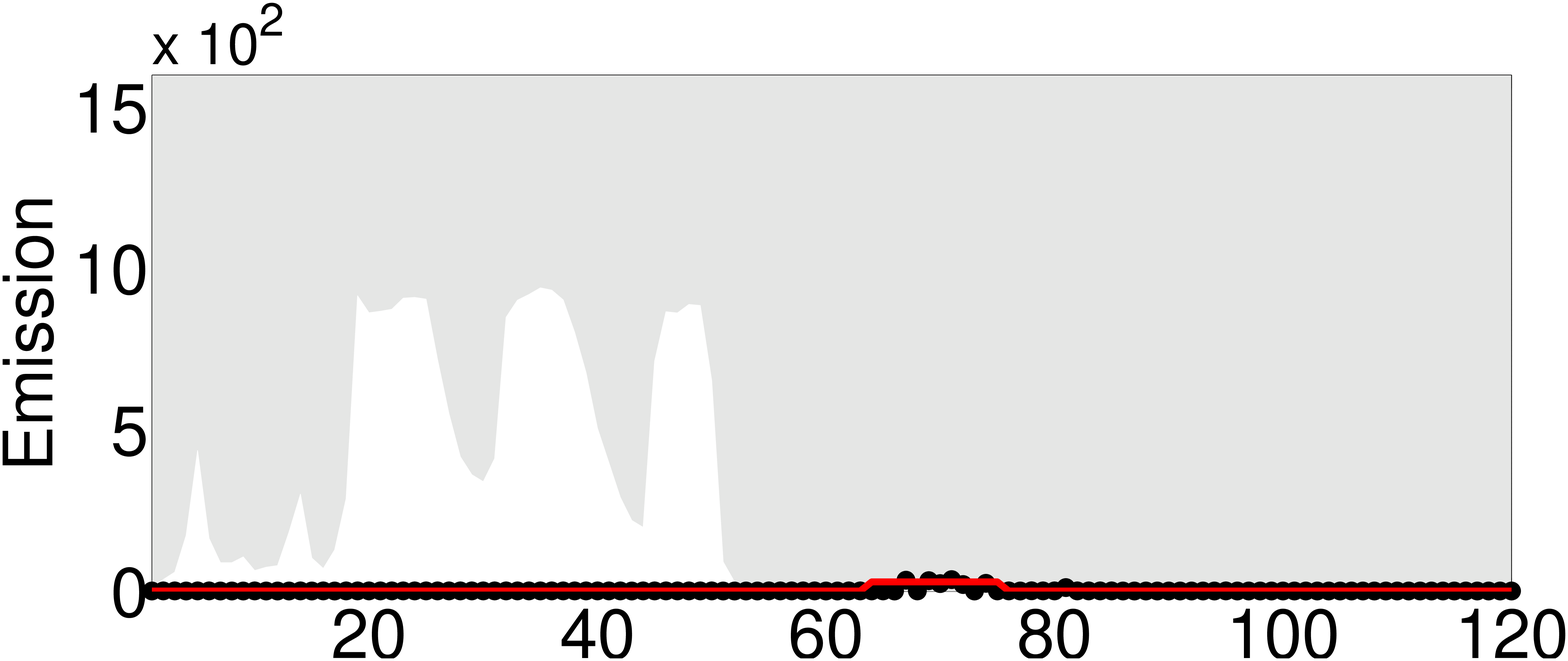}\vspace{-0.3cm}
 &
\includegraphics[width=1\linewidth]{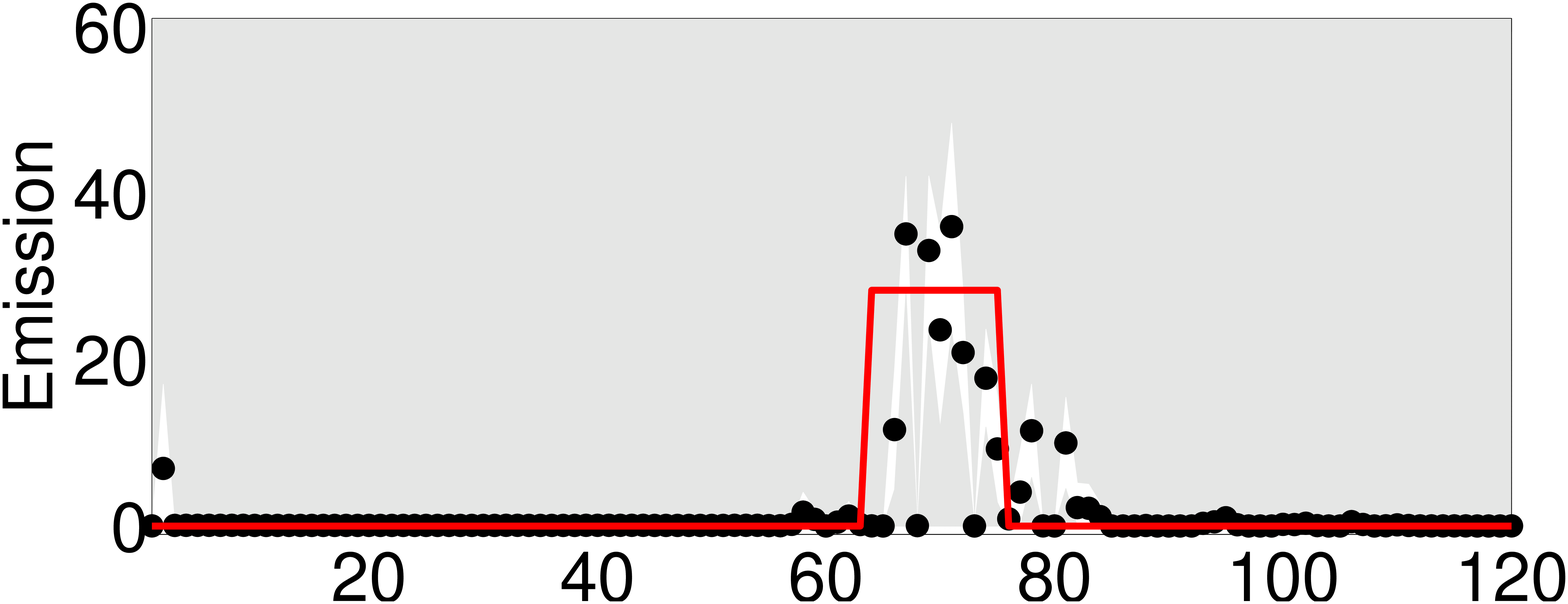}\vspace{-0.3cm}
\tabularnewline
\end{tabular}
\par\end{centering}
\caption{\label{fig:ETEX}Posterior distribution of the release profile for
the ETEX data set via its maximum and 95\% quantile. In all pictures,
the red line marks the ground truth for $\beta$, black circles are
the MAP estimate of $\beta$ and white area is the 95\% quantile.
The results of LS-APC method are accompanied by estimate of the correlation
parameter $l$ in the lower plot.}
\end{figure}

The resulting posterior distribution for the release profile are displayed
in Figure \ref{fig:ETEX} for the full ETEX data and in Figure \ref{fig:ETEX66}
for the ETEX 66. The reason is that the 95\% quantile of the release
profile is very wide at time indexes with extremely low sensitivity
of the measurements to the parameter as indicated in Figure \ref{fig:Visualization-of-matrix}.
The estimates at the time of the true release are almost identical
for both data sets as indicated by values of the absolute error of
the MAP estimate.

Note that all methods provide acceptable results given how uncertain
is the regressor matrix. It is impossible to draw definite conclusion
from comparison of the the posterior estimates with the true release.
Therefore, we will comment only on the qualitative indicators. 

First, we focus on description of the period of informative data which
is at the time of the true release. Note that the estimates of the
release provided by the Fused Lasso and Bayesian Fused Lasso are very
smooth with characteristic piece-wise constant shape. The uncertainty
of the estimation in the BLF is lower compared to that of the LS-APC
(GS) algorithm. On the other hand, the LS-APC algorithm enforces smoothness
only mildly, as demonstrated in comparison with results of the same
method with sparsity prior only (\ref{eq:ard}) which corresponds
to LS-AC with $l=0$. Results of the VB and GS approximations of the
LS-APC model are well comparable, with VB providing narrower uncertainty
bounds. 

\begin{table}
\begin{centering}
\begin{tabular}{ccccccc}
\toprule 
 &
FL &
BFL &
LS-APC &
LS-APC  &
LS-APC &
LS-APC\tabularnewline
 &
 &
 &
(GS) &
(VB) &
(GS) $l=0$ &
(VB) $l=0$\tabularnewline
\midrule 
ETEX &
199.95 &
233.80 &
181.49 &
180.85 &
231.19 &
230.48\tabularnewline
\midrule 
ETEX 66 &
194.94 &
215.98 &
186.52 &
180.85 &
224.34 &
231.34\tabularnewline
\bottomrule
\end{tabular}
\par\end{centering}
\caption{Absolute error of the MAP estimate of the release profile from the
true release profile of the ETEX experiment.}
\end{table}

However, the difference in uncertainty handling between the GS and
VB approximations is most visible at the period of uninformative data
which is for time index less than 55. In this period, the 95\% uncertainty
interval for the GS algorithm is extremely wide while for the VB approximation,
the posterior density is concentrated around zero. This is perhaps
the most interesting feature LS-APC (GS) method. Due to low sensitivity
of the measurements to the parameter values in this region, we can
not reliably conclude that the tracer was not being released at this
time. The LS-APC(GS) algorithms thus provides the most conservative
answer which is desirable in this application domain.

\section{Conclusion}

We have analyzed two prior models encouraging smoothness and sparsity
of the linear regression model, the Bayesian Fused Lasso and the Least
Squares with adaptive prior covariance matrix. The derived Gibbs sampling
algorithm for the latter was found to provide the best results on
simulated and real data from the European tracer experiment. However,
even the original Variational Bayes inference of the LS-APC model
was found to be well suitable for large data set arising in atmospheric
science due to its computational speed. The drawback of the variational
method is underestimation of the uncertainty of the estimate. However,
the variational lower bound for the marginal likelihood (i.e. Bayes
factor for the model selection problem) was found to be in very good
agreement with the same value provided by the Gibbs sampling. 

Since the LS-APC method allows to define an arbitrary conditionally
independent structure of the covariance matrix, the method can be
readily used in wide range of applications.

\bibliographystyle{23_home_smidl_work_STRADI_12_Users_ulrych_Bayesian_Analysis_2016_smj}
\bibliography{22_home_smidl_work_STRADI_12_Users_ulrych_Bayesian_Analysis_2016_bayesian,vs-world,ot}

\end{document}